\documentclass[11pt]{article} % use larger type; default would be 10pt
\pdfoutput=1 % if your are submitting a pdflatex (i.e. if you have
\usepackage{jheppub} % for details on the use of the package, please
\usepackage[utf8]{inputenc}
\usepackage[english]{babel}
\usepackage{amsmath,amssymb,amsbsy,amstext,amsthm,booktabs,simplewick,exscale,relsize,slashed,graphicx,amsfonts,upgreek, xcolor,subcaption}
\usepackage{multirow,color,dcolumn,bm,enumerate}
\usepackage{hyperref}
\usepackage{feynmp-auto,axodraw2,tikz}
\usepackage{ulem}
\usepackage{comment}
\usepackage{amsmath}
\usepackage{ytableau}
\usepackage{hyperref}
\usepackage[capitalise,noabbrev,nameinlink]{cleveref}

\title{Hidden $U(N)$ Symmetry behind $\mathcal{N}=1$ Superamplitudes}
\author[a]{Antonio Delgado,}
\author[a]{Adam Martin,}
\author[a]{Runqing Wang}

\affiliation[a]{Department of Physics, University of Notre Dame,
  South Bend, IN, 46556 USA}

\emailAdd{adelgad2@nd.edu}
\emailAdd{amarti41@nd.edu}
\emailAdd{rwang7@nd.edu}

\abstract{In this paper we develop a Young diagram approach to constructing higher dimensional operators formed from massless superfields and their superderivatives in $\mathcal{N}=1$ supersymmetry. These operators are in one-to-one correspondence with non-factorizable terms in on-shell superamplitudes, which can be studied with massless spinor helicity techniques. By relating all spin-helicity variables to certain representations under a hidden $U(N)$ symmetry behind the theory, we show each non-factorizable superamplitude can be identified with a specific Young tableau. The desired tableau is picked out of a more general set of $U(N)$ tensor products by enforcing the supersymmetric Ward identities. We then relate these Young tableaux to higher dimensional superfield operators and list the rules to read operators directly from Young tableau. Using this method, we present several illustrative examples.
}

\begin{document}
\maketitle

\setcounter{page}{2}

\section{Introduction}
\label{intro}

Effective field theories provide a framework to describe Nature in a particular energy range. One writes the most general Lagrangian (or Hamiltonian in the non-relativistic regime) consistent with the symmetries of the problem to address. The prescription is to include all possible operators to an arbitrary mass dimension built with fields and derivatives. The task is gigantic, as the number of operators increases with the mass dimension, and one needs to know how many independent terms exist.

 The goal of this paper is to develop a Young tableau (YT) approach \cite{Henning:2019mcv,Henning:2019enq,Li:2020gnx} to counting and determining the form of higher dimensional operators in $\mathcal N=1$\footnote{We use $\mathcal{N}$ to represent the number of supercharges to distinguish from the N in U(N), which represents the number of superfields.} supersymmetry. This approach is complementary to previous studies \cite{Delgado:2022bho,Delgado:2023ivp} where Hilbert series techniques were shown to work in $\mathcal N=1$ supersymmetric EFTs to determine the number of independent operators for a given mass dimension.
 
While superamplitudes have been explored extensively in the literatures \cite{Elvang:2010xn,Elvang:2009wd,Elvang:2011fx,Drummond:2008vq,Herderschee:2019ofc,Boels:2011zz,Elvang:2013cua}, these studies have been limited to renormalizable theories rather than effective field theories (theories containing higher dimensional, yet supersymmetric operators). Our starting point is the one-to-one correspondence between higher dimensional operators in non-supersymmetric theories and non-factorizable terms in on-shell amplitudes \cite{Arkani-Hamed:2017jhn,Cohen:2010mi,Arkani-Hamed:2008owk}. As this holds for component fields (scalars, fermions, gauge fields), extending the Poincare symmetry to supersymmetry requires finding possible correspondences between on-shell superamplitudes and higher dimensional operators formed from superfields and superderivatives. We will show that those relations do exist and therefore we can extend the use of Young tableaux to describe supersymmetric EFTs.  In this paper we will focus solely on operators formed from massless, distinguishable superfields. While we will show how to select the allowed Young tableau, we will not work out how to form a basis from all operators with the same YT shape. This detail will be presented in a companion paper~\cite{spinhelicitypaper2}, along with work extending the Young tableau superamplitude technique to indistinguishable superfields. 

Using on-shell amplitudes to describe EFTs~\cite{Shadmi:2018xan,Ma:2019gtx,Aoude:2019tzn,Durieux:2020gip,DeAngelis:2022qco,AccettulliHuber:2021uoa,Liu:2023jbq} allows one to use spinor-helicity techniques, which in turn dramatically simplifies redundancies from the equation of motions (EOM) and integration by parts (IBP), as they became relations in momentum space rather than in a more abstract field space. It also simplifies the explicit construction of the operators (e.g. going from the field content to exactly how all indices are contracted), as each interpolating field in an operator is replaced with spinor helicity variables that carry the appropriate Lorentz transformation and little group scaling properties. For example, fermions $\psi, \psi^\dag$ and field strengths $X$ are replaced by (taking all fields to be massless)
\begin{align}
\label{eq:replaceonshell}
\psi_{i,1/2} \to \lambda_{i,\alpha}, \quad \psi^\dag_{i,-1/2} \to \tilde \lambda^{\dot\alpha}_i, \quad X_{i, +1} \to \lambda_{i,\alpha}\lambda_{i,\beta}, \quad X_{i, -1} \to \tilde\lambda^{\dot\alpha}_i\tilde\lambda^{\dot\beta}_i,
\end{align}
where subscript $i,h$ labels the particle index and helicity respectively, and $\alpha, \dot\alpha$ are $SU(2)$ indices for the Lorentz group. Combined with $D_{i,\mu} \to \lambda_i \tilde\lambda_i = p_{i,\mu}$, the usual momentum replacement for on-shell momenta, the process of constructing the operators boils down to determining all possible spinor products.

While EOM redundancies are trivial in the spinor-helicity representation, e.g. $\slashed D\psi_i \to \tilde\lambda^{\dot\alpha}_i \lambda_{i,\alpha}\lambda^{\alpha}_i =  \tilde\lambda^{\dot\alpha}_i [\lambda_i \lambda_i] = 0 $, some manual manipulation is still needed to enforce IBP and sort through the allowed spinor products\footnote{We use the following conventions for spinor helicity variables (spinorial indices are suppressed unless necessary) $\lambda_{i,\alpha} \leftrightarrow |i], \quad \tilde \lambda^{\dot\alpha}_i \leftrightarrow |i\rangle $,
with the former representing positive helicity and the latter representing negative. The on-shell momentum of line $i$ is $p_{i,\alpha\dot\alpha} = \lambda_{i\alpha}\tilde \lambda_{i,\dot\alpha}$, while the total momentum is the sum over all particles $i = 1\cdots N$ in the amplitude $P_{\alpha\dot{\alpha}}=\sum_{i=1}^N \lambda_{i\alpha}\tilde{\lambda}_{i\dot{\alpha}}$.}. However, in a further step, Ref.~\cite{Henning:2019enq,Henning:2019mcv,Li:2020gnx} exploited a hidden $U(N)$ symmetry, where $N$ is the number of fields in the amplitude/operator, under which the $\lambda/\tilde \lambda$ transform as fundamentals/anti-fundamentals. Products of $\lambda, \tilde\lambda$ (e.g. a non-factorizable amplitude) can be decomposed using Young tableau techniques, and the $U(N)$ organization allows one to immediately spot and remove IBP redundancies (a review of the basics of non-supersymmetric YT construction are given in Appendix \ref{Non-Supersymmetric SSYT}) . 

The rest of the paper is organized as follows. We first review the basics of the supersymmetry and on-shell supersymmetric states in Section 2. We then move, in Sections 3 and 4, to apply these concepts to higher dimensional $\mathcal N=1$ supersymmetry operators and on-shell superamplitudes. This will lead us to a  supersymmetry extension of Eq.~\eqref{eq:replaceonshell}  -- a rule for how to replace superfields and superderivatives in superamplitudes. The rules involve spinor-helicity variables supplemented by additional Grassmann variables that enforce supersymmetry relationships among component amplitudes of different helicity. Sections 5 contains the main result of the paper, the translation from superfield operators to Young tableau. We devote Sections 6 to our conclusions, while more some detailed calculations are presented in the appendices.

\section{Supersymmetry and On-shell supersymmetric states}\label{superstate}

The underlying algebra of supersymmetry is the super-Poincar\'e algebra, which extends the usual ten-parameter Poincar\'e algebra in four dimensions by adding $2\mathcal{N}$ fermionic generators $Q_{\alpha A},Q^\dag_{\dot{\alpha} B}$, where $\alpha,\dot{\alpha}$ are spinor indices and $A,B=1,2,\cdots \mathcal{N}$. For $\mathcal{N}=1$ supersymmetry, the anti-commutation relations (because supercharges $Q_\alpha,Q^\dag_{\dot{\alpha}}$ are fermionic) are defined as:
\begin{equation}\label{q algebra}
\begin{split}
\{ Q_\alpha,Q^\dag_{\dot{\beta}}\}&=-2(\sigma^\mu_{\alpha\dot{\beta}})P_\mu, \\
\{ Q_\alpha,Q_\beta\}&=0,\\
\{ Q^\dag_{\dot{\alpha}},Q^\dag_{\dot{\beta}}\}&=0.
\end{split}
\end{equation}

On-shell non-supersymmetric states are labeled by their momentum and spin, which means $ |s_i, p_i \rangle \to |s_i,\lambda_i, \tilde \lambda_i \rangle$. Supersymmetry relates states with different spin. Rather than working with the individual component states, it's convenient to group all states related by supersymmetry  into a single {\it superstate}. This can be accomplished via the coherent state formalism familiar from e.g. fermionic oscillators. We introduce a Grassmann variable $\eta_i$ for each on-shell supermultiplet in the amplitude and define the superstate as~\footnote{Our conventions here differ slightly from Ref.~\cite{Arkani-Hamed:2008owk}.}
\begin{align}
\label{eq:coherenteta}
|\eta_i, \lambda_i, \tilde \lambda_i \rangle \equiv e^{Q_{\alpha} \omega^{\alpha}\eta_i}|s_i, \lambda_i, \tilde \lambda_i \rangle
\end{align}
where $\omega$ is a spinor that satisfies $[\omega \lambda_i] = 1$. This definition assumes a convention where $Q_\alpha |-s \rangle =  Q^{\dag,\dot\alpha}|+s\rangle = 0$, and (suppressing $\lambda, \tilde \lambda$ labels in the states)
\begin{align}
Q_\alpha |s_i \rangle =  \lambda_{i\alpha} |s_i - 1/2 \rangle, \quad Q^{\dag,\dot\alpha}|-s_i \rangle = \tilde\lambda^{\dot\alpha}_i|-s_i + 1/2 \rangle, 
\end{align}
so that:
\begin{align}
\label{eq:etastate}
|\eta  \rangle = |s \rangle + \eta_i \omega^{\alpha} Q_{\alpha}  |s \rangle  = |s \rangle + \eta_i \omega^{\alpha}  \lambda_{\alpha}  |s-1/2  \rangle = |s \rangle + \eta_i |s-1/2 \rangle.
\end{align}

The $|\eta \rangle$ are, therefore, eigenstates of $Q^\dag$:
\begin{align} 
Q^{\dag,\dot\alpha} |\eta_i \rangle = Q^{\dag,\dot\alpha} ( |s \rangle + \eta_i |s-1/2\rangle ) = \eta_i Q^{\dag,\dot\alpha} |s-1/2\rangle  =  \eta_i \tilde \lambda^{\dot\alpha} |s \rangle  =  \eta_i \tilde \lambda^{\dot\alpha} (|s \rangle + \eta_i |s-1/2 \rangle) = \eta_i \tilde \lambda^{\dot\alpha} |\eta_i \rangle, \nonumber
\end{align}
where we have used $\eta^2_i = 0$. As such, the $|\eta_i \rangle$ are simply re-phased under $Q^\dag$ supersymmetry transformations,
\begin{align}
e^{Q^{\dag\dot\alpha} \tilde \xi_{\dot\alpha}}|\eta_i \rangle = e^{\eta_i \langle \tilde\xi\tilde\lambda \rangle}|\eta_i \rangle
\end{align}
Under $Q$ supersymmetries, $\eta_i$ shifts
\begin{align}
e^{Q_{\alpha} \xi^\alpha} |\eta_i \rangle &= e^{Q_{\alpha} \xi^\alpha + Q_{\alpha} \omega^{\alpha}\eta_i}|s_i \rangle  \nonumber \\
& = (1 +  Q_{\alpha}(\xi^\alpha + \omega^{\alpha}\eta_i)|s_i\rangle = |s_i \rangle + (\xi^{\alpha} + \omega^{\alpha}\eta_i)\lambda_{\alpha}|s_i-1/2 \rangle \nonumber \\
& = |s_i \rangle + (\eta_i +  [ \xi  \lambda ]) |s_i-1/2 \rangle = |\eta_i + [  \xi  \lambda_i ] \rangle
\end{align}

This behavior -- re-phasing under $Q^\dag$ and translation in $\eta$ under $Q$ -- is faithfully captured if we make the identification
\begin{align}
 \quad Q_i^{\dag,\dot\alpha} = \tilde \lambda^{\dot\alpha}_i \eta_i , \quad Q_{i, \alpha} = \lambda_{i\alpha} \frac{\partial}{\partial \eta_i} 
\end{align}
where the $i$ index indicates we have projected the supercharge along the momentum of on-shell particle $i$\footnote{The projected supercharges are often referred to as $q^\dag, q$ in the literature.}. For an amplitude consisting of $N$ superstates, the total supercharges are the sum over the individual projections, 
\begin{equation}\label{q rep}
Q^{\dag,\dot\alpha} =  \sum_{i=1}^N  \tilde \lambda^{\dot\alpha}_i \eta_i \quad \quad Q_\alpha = \sum_{i=1}^N \lambda_{i\alpha} \frac{\partial}{\partial \eta_i}
\end{equation}

Acting with $Q^\dag$ supersymmetry on an amplitude $A(\eta_i)$ of superstates, $A(\eta_i)$ only changes by an overall phase
\begin{align}
Q^\dag\,A(\eta_i) = e^{(\sum_i \tilde \lambda_i \eta_i)\tilde \xi}A(\eta_i),
\end{align} 
Something analogous happens under translation (momentum generator) on a non-supersymmetric amplitude $AA(p_i)$
\begin{align}
P\, A(p_i) = e^{ix \cdot \sum_i p_i} A(p_i),
\end{align} 
from which we conclude $A(p_i)$ must be proportional to a total momentum conserving delta function $A(p_i) \propto \delta(\sum_i p_i)$. The identical logic for the supersymmetry case tells us $A(\eta_i) \propto \delta^{(2)}(\sum_i \tilde \lambda_i \eta_i) = \delta^{(2)}(Q^\dag)$.

We will work with the convention that a chiral superfield $\Phi$ contains a helicity $+1/2$ fermion $\psi$ and a scalar, while $\Phi^\dag$ contains the complex conjugate scalar and a helicity $-1/2$ fermion $\psi^\dag$. Using Eq.~\eqref{eq:etastate} and identifying the component states $\Phi, \Phi^\dag$ create, we have
\begin{align}
\label{eq:swavefn}
\Phi_i = \psi_i + \eta_i\, \phi_i, \nonumber \\
\Phi^\dag_i = \phi^*_i + \eta_i\, \psi^\dag_i
\end{align}
The identifications above should be viewed as superwavefunctions for on-shell amplitudes, and not confused with the off-shell expansion of superfields into components fields, $\theta, \bar\theta$. On-shell amplitudes are polynomials of the superwavefunctions, and expanding the wavefunctions as above allows one to pick out component subamplitudes (by differentiating with respect to various $\eta_i$ or setting $\eta_i \to 0$)\footnote{ See Sect.~\ref{On-Shell Versus Off-Shell} for more detail on the superwavefunctions and their relation to off-shell supersymmetry.}.

Another enormous benefit of spinor-helicity variables is that they make an amplitude's little group properties manifest. For massless particles, the little group transformations are $\lambda_i \to t_i \lambda_i, \tilde \lambda_i \to t^{-1}_i \tilde \lambda_i$; here $t_i$ is a phase (as the massless little group is $U(1)$) and note that $p^\mu_i$ is left invariant. The overall little group scaling of an amplitude is connected to the helicities of the particles involved. To make the supercharges individually little group invariant, we extend the little group properties to $\eta_i$, $\eta_i \to t_i\, \eta_i, \partial/\partial \eta_i \to t_i^{-1}\partial/\partial \eta_i$. 

In Eq.~\eqref{eq:coherenteta}, we made the choice to identify superstates by starting with the highest helicity component ($|+1/2 \rangle$ for a chiral superfield) and acting with $Q^\dag$ to create the $|0\rangle$ state. This choice fixed how the different supersymmetry generators acted on the states. However, this is not the only possibility. We could have expressed the same superstate by starting with the lowest helicity component and acted with $Q$ to create the $|+1/2\rangle$. Done this way, the expression analogous to Eq.~\eqref{eq:coherenteta} is
\begin{align}
e^{Q^{\dag,\dot\alpha} \tilde\omega_{\dot\alpha} \bar\eta_i}|s_i, \lambda_i, \tilde\lambda_i \rangle \equiv |\bar\eta_i, \lambda_i, \tilde\lambda_i \rangle
\end{align}
where $\langle \tilde \omega \tilde \lambda \rangle = 1$ and $\bar\eta$ is another Grassmann variable unrelated to $\eta$.
The $|\bar\eta \rangle$ superstate representations are eigenstates of $Q$. Following the same steps as before, we see the action of $Q$ and $Q^\dag$ supersymmetry are reversed on $|\bar \eta \rangle$ compared to $|\eta\rangle$ -- $Q^\dag$ supersymmetry translates $\bar \eta$ while $Q$ supersymmetry rephases $|\bar \eta \rangle$. The latter property implies that the action of $Q$ on an amplitude $A(\bar \eta)$ of $\bar \eta$ superstates is an overall phase, and thus $A(\bar\eta) \propto \delta^{(2)}(Q)$.

The fact that there are two different representations of the same superstate (and therefore, of any product of superstates) can be confusing at first, as it changes how supercharges behave, but is actually quite a powerful tool. The two different representations -- or `bases' -- can be converted into each other using a Grassmann Fourier transform. 
\begin{align}
|\bar\eta_i \rangle = \int d\eta_i\, e^{\eta_i\bar\eta_i}|\eta_i \rangle, \quad \quad |\eta_i \rangle = \int d\bar\eta_i\, e^{\bar \eta_i\eta_i}|\bar \eta_i \rangle
\end{align}
 For the bulk of this paper we will use the $\eta$ basis whenever we can, though as explained in Sec.~\ref{unYT} there are some instances where this is not convenient and we will need to switch to the $\bar\eta$ form.

\section{Superamplitudes}
\label{superamplitudes}

A superfield operator is defined as an invariant under supersymmetry transformations. It's well known that there are two ways to construct such invariants: integrating a holomorphiomic superpotential $W(\Phi)$ over half of the superspace or integrating a real K\"ahler potential $K(\Phi,\Phi^\dagger)$ over the entire superspace, denoted as $F$-term $(W_F)$ and $D$-term $(K_D)$ respectively:
\begin{equation}\label{W,K}
W_F\equiv\int d^2\theta W(\Phi),\ \ \ \ \ K_D\equiv\int d^4\theta K(\Phi,\Phi^\dagger),
\end{equation}

 For example, the renormalizable Lagrangian for a single chiral superfield $\Phi$ is given by:
\begin{equation}\label{lagrangian}
\mathcal{L}=(W_F+h.c.)+K_D=\int d^2 \theta[(\frac{1}{2}m\Phi^2+\frac{1}{3}g\Phi^3)+h.c.]+\int d^4\theta(\Phi\Phi^\dagger),
\end{equation}
where the two functions are chosen to be $W(\Phi)=\frac{1}{2}m\Phi^2+\frac{1}{3}g\Phi^3$ and $K(\Phi,\Phi^\dagger)=\Phi\Phi^\dagger$.

Furthermore, as shown in \cite{Delgado:2022bho}, any superpotential term containing superderivatives can be written as a linear combination of K\"ahler terms. As most of the complications in building superamplitudes will come from superderivatives, we will focus on K\"ahler operators for the majority of this work. A discussion of how higher dimensional (zero derivative) superpotential operators fit into our method can be found in Appendix \ref{F term}.

In theories with extended $\mathcal N > 1$ supersymmetry, $R$-symmetries help constrain the form of the superamplitudes, so it is natural to ask whether that is the case for our study. We find that it is not. This stems from the fact that $\mathcal N = 1$ supersymmetry permits a much wider set of interactions than are allowed if $\mathcal N > 1$. In fact, it is entirely possible that an $\mathcal N = 1$ theory with sufficiently many interactions (for a given field content) does not even respect an $R$ symmetry at all.

\subsection{Superamplitudes: Examples}\label{example}

To get a better understanding of superamplitudes, let us turn to some simple examples. We explore these from the bottom up, meaning we explicitly expand the superfields into components, then explore the component subamplitudes with spinor helicity variables. As we will show, the connection between subamplitudes required by supersymmetry will be maintained by adding Grassmann weights to certain terms. Using these examples, we will use what we have learned to generalize to other operators and provide a more direct approach using a `replacement rule' along the lines of Eq.~\eqref{eq:replaceonshell}.

We start with the simplest example $\Phi_1\Phi_2\Phi^\dag_3\Phi^\dag_4$\footnote{This operator is in principle not real, but one can always adds its hermitian conjugate to make the entire operator real, and thus becomes a valid $D$-term (K\"ahler term).}, an operator with two chiral superfields, two anti-chiral superfields, and no derivatives. Here we've written this using off-shell superfields (so, with $\theta$ dependence) rather than superstate Eq.~\eqref{eq:swavefn}.  We will take all fields to be distinguishable and massless. Let's expand this operator, taking
\begin{align}
\Phi_i \to \phi_i + \sqrt 2 \theta \psi_i  +i\, \theta \sigma^\mu \bar\theta \partial_\mu \phi, \quad \Phi^{\dag}_i \to \phi^*_i + \sqrt 2 \bar\theta \psi^\dag_i  - i\,\theta \sigma^\mu \bar\theta \partial_\mu \phi^*. 
\end{align}
Note that the pieces with higher powers of $\theta, \bar\theta$ -- $\theta^2 \bar\theta^2\Box \phi_i, \theta \theta \sigma^\mu \bar\theta \partial_\mu \psi_i$  and $\theta^2 F$ -- all involve the EOM, and since we are taking all fields to be massless we have dropped these pieces. In doing this, we've truncated the off-shell superfield expansions to the pieces that survive on-shell.

Doing the $d^4\theta$ integral we get the component Lagrangian: 
\begin{equation}
\begin{split}
\int d^4 \theta (\Phi_1\Phi_2\Phi^\dag_3\Phi^\dag_4)&\supset  \partial\phi_1\partial\phi_2\phi_3^*\phi_4^*+\psi_1\partial\phi_2\psi_3^\dag\phi_4^*+\psi_1\partial\phi_2\phi_3^*\psi_4^\dag\\&+\partial\phi_1\psi_2\psi_3^\dag\phi_4^*+\partial\phi_1\psi_2\phi_3^*\psi_4^\dag+\psi_1\psi_2\psi_3^\dag\psi_4^\dag.
\end{split}
\end{equation}
Each term on the left hand side is a valid (higher dimensional) non-supersymmetric Lagrangian term, and therefore corresponds to an amplitude expressible in terms of spinor helicity variables. To go from the component amplitudes to the full superamplitude from $\Phi_1\Phi_2\Phi^\dag_3\Phi^\dag_4$, we need to add the proper Grassmann weights ($\eta_i$). The way to add proper weights is dictated by the coherent state picture of the on-shell superwavefunctions in Eq.~\eqref{eq:swavefn}. Specifically, for each field $\Phi_i$ in an amplitude, we include an $\eta_i$ when picking out the $\phi$ component (only), while for each $\Phi^\dag_i$ we include an $\eta_i$ when picking out the $\psi^\dag_i$ component. Performing these steps on the previous result, we have: 
\begin{equation}
\begin{split}\label{eg1}
A(\Phi_1\Phi_2\Phi^\dag_3\Phi^\dag_4)&=[12]\langle12\rangle\eta_1\eta_2+[12]\langle23\rangle\eta_2\eta_3+[12]\langle24\rangle\eta_2\eta_4\\
&+[12]\langle13\rangle\eta_1\eta_3+[12]\langle14\rangle\eta_1\eta_4+[12]\langle34\rangle\eta_3\eta_4\\
&=[12]\sum_{i<j}^4 \langle ij\rangle\eta_i\eta_j,
\end{split}
\end{equation}
Note that, given this full superamplitude, we can derive the component amplitudes by differentiating with respect to corresponding weights~\cite{Elvang:2011fx}.

For our second example, let's look at $\overline{D}D\Phi_1D\Phi_2\overline{D}\Phi^\dag_3$. We repeat the same steps as in the previous example, first expanding the fields into on-shell components, dropping EOM pieces, then integrating over $d^4\theta$:  
\begin{equation}
\begin{split}
\int d^4 \theta (\overline{D}D\Phi_1D\Phi_2\overline{D}\Phi^\dag_3)&\supset  \partial_{\{\mu,\nu\}}\phi_1\partial^\mu\phi_2\partial^\nu\phi_3^*+\partial_{\{\mu,\nu\}}\phi_1\psi_2\sigma^\mu \partial^\nu \psi_3^{\dag}+\partial_\mu \psi_1 \partial_\nu \phi_2 \sigma^\mu \partial^\nu\psi_3^\dag
\end{split}
\end{equation}
where $\partial_{\{\mu,\nu\}}$ is the symmetric combination of two derivatives (as $\Box \phi$ is removed by EOM).

Identifying the on-shell superfields with components of the superstates and expressing things  in terms of spinor-helicity variables, this becomes: 
\begin{align}
\label{eg2}
A(\overline{D}D\Phi_1D\Phi_2\overline{D}\Phi^\dag_3)&=[12]\langle13\rangle[13]\langle12\rangle\eta_1\eta_2+[12]\langle13\rangle[13]\langle13\rangle\eta_1\eta_3+[12]\langle13\rangle[13]\langle 23 \rangle\eta_2\eta_3\nonumber \\
&=[12]\langle13\rangle[13]\sum_{i<j}^3 \langle ij\rangle\eta_i\eta_j. 
\end{align}

We now see that both expressions Eqs.~\eqref{eg1},\eqref{eg2} contain the second-order Grassmann delta function:
\begin{equation}\label{delta1}
\delta^2({Q^\dag})\equiv\sum_{i<j}^N \langle ij\rangle\eta_i\eta_j,
\end{equation}
where $N$ is the number of total superfields. This is exactly the property for superamplitude in the $\eta$ basis explained in previous section. Furthermore, we will show in next section that the existence of this special function is not merely a coincidence but has a deep connection with super-Ward identities.

\subsection{Super-Ward Identities}\label{sec:superward}

The most general structure for a superamplitude must take into account how it transforms with respect to supersymmetry. Namely, any
superamplitudes must satisfy super-Ward identities, i.e. super-momentum conservation. For $\mathcal N=1$ supersymmetry, there are two sets of  conservation laws originating from four supercharges $Q_\alpha$ and $Q^\dag_{\dot{\alpha}}$, where $\alpha,\dot{\alpha}=1,2$, any amplitude $A$ must satisfy:
\begin{subequations}
\begin{equation}\label{QA}
Q_\alpha A=0,
\end{equation}
\begin{equation}\label{QdA}
Q^\dag_{\dot{\alpha}}A=0.
\end{equation}
\end{subequations}

To find a solution to these two constraints, recall that the action of $Q^{\dag}_{\dot{\alpha}}$ on coherent states in the $\eta$ basis implies
\begin{equation}
\label{eq:aeta}
A \propto \delta^2(Q^\dag)
\end{equation}
(as we saw in the examples of the previous section). Viewed in light of the super-Ward identities, this form automatically satisfies Eq.~\eqref{QdA}. The fact that the amplitude is proportional to a delta function should not be a surprise. In the case of  non-supersymmetric amplitudes, total momentum conservation:
\begin{equation}
P_\mu A=0.
\end{equation}
is satisfied by taking $A \propto \delta^4(P)$, where $\delta^4(P)$ is the usual 4-momentum delta function.

Notice that \eqref{eq:aeta} is only realized under in the $\eta$ basis for coherent states. If one goes to $\overline{\eta}$ basis, then the superamplitude satisfies:
\begin{equation}
A \propto \delta^2(Q),
\end{equation}
where $\delta^2(Q)$ is given by
\begin{equation}
\delta^2(Q)\equiv \sum_{i<j}^N [ij]\eta_i\eta_j.
\end{equation}

While the delta function takes care of one Ward identity, we still need to consider the second Ward identity (Eq.\eqref{QdA} in the $\eta$ basis). From our discussion in Sec. ~\ref{superstate}, we know that  $Q$ does not act simply on $\eta$ basis amplitudes, so the resolution to Eq.\eqref{QdA} must be more subtle than simply having the amplitude be proportional to $\delta^2(Q)$. 

As an initial step, let us verify that our two examples Eq.~\eqref{eg1}, \eqref{eg2} satisfy $Q_\alpha A = 0$. This is straightforward to see using $Q=\sum \lambda \frac{\partial}{\partial\eta}$, as the only $\eta$ dependence in Eq.~\eqref{eg1}, \eqref{eg2} lies in $\delta^{(2)}(Q^\dag)$, and $Q\delta^{(2)}(Q^\dag) = 0$ once we impose momentum conservation\footnote{$Q\,\delta^{(2)}(Q^\dag) = \sum_k \lambda_k \partial/\partial \eta_k (\sum\limits_{i<j}^n \tilde \lambda_i \tilde \lambda_j \eta_i\eta_j) = (\sum_k  \lambda_k \tilde \lambda_k)\, \sum_i \tilde \lambda_i \eta_i = 0$.}.

 To go further and see how to enforce $Q_\alpha A = 0$ on more general amplitudes, one hint comes from amplitudes in $\mathcal N = 1$ super Yang Mills theory, which can be determined from $\mathcal N = 4$ super Yang Mills amplitudes by `zeroing' the $\eta$ coordinates corresponding to the extraneous 3 supersymmetries~\cite{Elvang:2009wd}. There, one finds the structure 
\begin{equation}
\label{eq:mabc}
A \propto m_{abc},\quad m_{abc}\equiv \lambda_a\lambda_b\eta_c+\lambda_b\lambda_c\eta_a+\lambda_c\lambda_a\eta_b.
\end{equation}
which is shown to satisfy Eq.~\eqref{QA}. We cannot immediately just grab $m_{abc}$ for our purposes, as it comes from amplitudes in a purely renormalizable theory ($\mathcal N = 1$ super Yang Mills), while we are interested in non-factorizable amplitudes/higher dimensional operators, and it is limited to only three arguments. To proceed, we will first introduce `replacement rule' which allows us to directly express off-shell superfield operators directly in terms of spinor helicity variables. With this shortcut in hand, we'll then find the generalization of Eq.~\eqref{eq:mabc} using a diagrammatic method .

\section{Path from Off-shell to On-shell}
Let's summarize the facts we know about superamplitudes:
\begin{itemize}
\item Superamplitudes are the sum of all component amplitudes, weighted with Grassmann variables determined by the basis choice ($\eta$ vs. $\bar \eta$) and what superfields are present.
\item super-Ward identities put strong constraints on the form of superamplitudes. We can express these constraints in terms of spin helicity variables.
\end{itemize}
Recall that in non-supersymmetric case Eq.~\eqref{eq:replaceonshell} relates spin-helicity variables with asymptotic fields. The theory is then solved in its on-shell form under Ward identities constraint, and the off-shell result can be restored using Eq.~\eqref{eq:replaceonshell}.  A direct generalization of Eq.~\eqref{eq:replaceonshell} to superfields is not straight forward because supermultiplets contain more than one field (not represented by a single helicity). Nevertheless, using a combination of spinor helicity and coherent state Grassmann variables, we have found a bridge between on-shell and off-shell supersymmetric theory.

\subsection{Replacement Rules}

 As stated earlier, the one-to-one correspondence between on-shell amplitudes and operators only holds for higher dimensional operators, so we only need to concern ourselves with how higher dimensional operators in $W_F, K_D$ translate on-shell. As $W_F,K_D$ contain products of superfields, our first step is a translation rule for superfields, the analog of Eq.~\eqref{eq:replaceonshell}. For now, we will focus on (massless) chiral/anti-chiral superfields. An example using massless vector superfields can be found in Appendix~\ref{app:vector}. In addition to a translation/replacement rule for superfields, we also need an on-shell rule for the superderivatives $D_\alpha, \overline D^{\dot\alpha}$.

Consider the following replacement for chiral superfields,
\begin{equation}\label{field rep}
\Phi_i \rightarrow \eta_i,\ \ \ \Phi_i^\dag\rightarrow 1,
\end{equation}
coupled with the rule
\begin{equation}\label{d rep}
D_{i,\alpha}= \lambda_{i\alpha} \frac{\partial}{\partial\eta_i},\ \ \ \ 
\overline{D}^{\dot{\alpha}}_i=\tilde{\lambda}_{i\dot{\alpha}}\eta_i,
\end{equation}
for superderivatives. As in Eq.~\eqref{eq:replaceonshell}, the replacements represent the Lorentz properties of the objects ($\Phi, \Phi^\dag$ are scalars while $D, \overline D$ are spinors), but there are some differences. First, the replacements of $\Phi, \Phi^\dag$ only seem to capture the lowest (in magnitude) helicity components of the superfield. Second, the replacements look asymmetric, as $\Phi$ is replaced with a Grassmann variable while $\Phi^\dag$ is not. This asymmetry is linked to our use of the $\eta$ basis for states, where the $\eta$ accompanies the lowest helicity component of the superstate. Had we defined $\Phi$ superstates with the $\eta$ basis but used the $\bar\eta$ basis for $\Phi^\dag$ states, the replacement would look more symmetric, but this would make later steps more difficult. The replacement $\Phi_i \to \eta_i$ may look worrisome, as it seems to imply that $\Phi^2_i \to 0$. For now, we will focus on operators/amplitudes formed from distinguishable fields, where this issue does not arise. The proper treatment of indistinguishable fields will be presented in a companion paper~\footnote{As we will show in a forthcoming work, to avoid $\Phi^2 = 0$ for identical superfields we need to add an extra index on $\eta$ for identical superfields.}. 

To justify the rules in Eqs.~\eqref{field rep}, \eqref{d rep}, let us carry out $D\Phi, \overline D\Phi$, etc.
\begin{align}
\label{eq:Dphistuff}
D_{i,\alpha}\Phi_i = \lambda_{i\alpha} \frac{\partial}{\partial\eta_i}(\eta_i) = \lambda_{i,\alpha} \quad & \quad \overline D_{i,\dot\alpha} \Phi_i = \tilde{\lambda}_{i\dot{\alpha}}\eta_i \eta_i = 0  \\
D_{i,\alpha}\Phi^\dag_i = \lambda_{i\alpha} \frac{\partial}{\partial\eta_i}(1) = 0 \quad & \quad \overline D_{i,\dot\alpha} \Phi^\dag_i = \tilde{\lambda}_{i\dot{\alpha}}\eta_i (1)= \tilde{\lambda}_{i\dot{\alpha}}\eta_i.
\end{align}
We see that the usual annihilation conditions of chiral/anti-chiral superfields are now realized thanks to the Grassmann nature of $\eta_i$. Further, $D\Phi, \overline D\Phi^\dag$ replacements carry the Lorentz properties of the fermionic (highest helicity) components. Finally, the replacements automatically remove EOM redundancy, as 
\begin{align}
\label{eq:repEOM}
D^2_i \Phi_i = \lambda_{i\alpha} \frac{\partial}{\partial\eta_i}(\lambda_{i, \alpha}) = 0 \quad & \overline D^2_i\Phi^\dag_i = \tilde \lambda_{i,\dot\alpha}\tilde \lambda^{\dot\alpha}_i \eta_i\eta_i  = \langle i i \rangle \eta_i\eta_i  = 0.
\end{align}

Following replacements Eqs.~\eqref{field rep} and \eqref{d rep}, we can translate any higher dimensional superfield operator into on-shell language. However, we are not finished because we still need to integrate over the superspace coorrdinates $d^2\theta, d^4\theta$ depending on whether the operator resides in the superpotential or K\"ahler potential. As Grassmann integration is identical to differentiation, these $d^2\theta, d^4\theta$ can be replaced with additional powers of $D, \overline D$:
\begin{align}
d^2\theta  \rightarrow D^2\Big|_{\theta = 0}, \quad d^4\theta \rightarrow \overline D^2 D^2\Big|_{\theta = \bar\theta = 0},
\end{align}
where the $D, \overline D$ appearing here are the {\it total} derivatives, meaning summed over all particles in the amplitude/operator: $D = \sum_i D_i, \overline D = \sum_i \overline D_i$. These `extra' derivatives play an vital role in enforcing the super-Ward identities.

One may have noticed that the replacements for $D_{i,\alpha}, \overline D^{\dot\alpha}_i$ are identical to the representation of the on-shell supercharges $Q_{i,\alpha}, Q^{\dag, \dot\alpha}_i$ developed in Sec.~\ref{superstate}. Given that $Q_{i,\alpha}, Q^{\dag, \dot\alpha}_i$ are generators of the supersymmetry algebra, while $D_{i,\alpha}, \overline D^{\dot\alpha}_i$ are not, it may seem fishy that they have the same on-shell replacement. Relatedly, Eqs.~\eqref{eq:Dphistuff}-\eqref{eq:repEOM} may seem suspect as they equip the superstate with chiral/EOM constraints that we usually associate with off-shell fields.  However, as we will discuss in next section, there is indeed a deep relation between superstates and superfields which justifies the above arguments. Following that discussion, we will return to superamplitude construction in Sec.~\ref{sec:unsymm}.

\subsection{Superfield versus Superstate}\label{On-Shell Versus Off-Shell}

The coherent states (Eq. \eqref{eq:swavefn}) we introduced are different from the off-shell chiral/anti-chiral superfield we use to build a Lagrangian (Eq.~\eqref{W,K}, \eqref{lagrangian}), although they are represented using the same symbol. The former is on-shell, and therefore has 1+1 (bosonic+fermionic) degrees of freedom, while the latter one has 1+2+1 degrees of freedom, where the last 1 comes from an auxiliary field. As such, our replacement rule (Eq.~\eqref{field rep}) may seem confusing, as it equates the two (off shell superfield with a coherent state). However, there is actually nothing wrong with this identification, because what we are really identifying is the lowest component of a superfield with a coherent superstate built from this component field. 

Recall that an off-shell superfield $S$ is generated by its lowest component field $A$ according to $S(x,\theta,\overline{\theta})=e^{(\theta Q+\overline{\theta}\overline{Q})} A(x)$, where $Q$ and $\overline{Q}$ are group generators (supercharges) of $\mathcal N=1$ supersymmetry. Similarly, an on-shell superstate is generated by its lowest component field, as introduced in Section \ref{superstate}. Therefore, if we can identify the component fields upon removing EOM (such that they have the same degrees of freedom), the related superfield operator and superstate are also identified, which allows us to study everything in a manifestly supersymmetric way.

For example, let's look at a coherent state $\Phi$ given by Eq.~\eqref{eq:swavefn}:
\begin{equation}
\Phi= \psi+ \eta\, \phi, \nonumber,
\end{equation}
and a superfield $\boldsymbol{\Phi}$ (in this section we will use a bold symbol to represent a superfield and a supercovariant derivative defined in superspace):
\begin{equation}\label{phi}
\boldsymbol{\Phi}=\phi(x)+\theta\psi(x)+\theta^2F(x).
\end{equation}
We are making the following identification using \eqref{field rep} in the first step:
\begin{equation}
\Phi \sim \eta\sim \Phi|_\eta\sim \phi\sim \boldsymbol{\Phi}|_{\substack{on-shell \\ \theta=\overline{\theta}=0}},
\end{equation}
where the LHS is the $\phi$ component that generates the entire on-shell coherent superstate while the RHS is the lowest component field which generates the entire off shell superfield. The same identification also works when we add supercharges/derivatives:
\begin{equation}
\label{QandD}
Q\Phi = \lambda \frac{\partial}{\partial\eta}(\eta) = \lambda\sim(Q\Phi)|_{without\ \eta}\sim \psi\sim \boldsymbol{D}\boldsymbol{\Phi}|_{\substack{on-shell \\ \theta=\overline{\theta}=0}}
\end{equation}
where the on-shell condition reduces the degrees of freedom in $\psi$ from two to one. The reason that $Q^2\Phi\sim0$ is manifest in this sense is because $\boldsymbol{D}^2\boldsymbol{\Phi}=0$ for massless chiral superfield upon using the equations of motion. Equation \eqref{QandD} also explains why $\boldsymbol{D}(\boldsymbol{\overline{D}})$ and $Q(Q^\dag)$ have the same spinor helicity form (up to a constant). For anti-chiral superfields everything works similarly, so we will not repeat the relations here. 

With this identification rule, it should be clear from the context which $\Phi$ we mean in this paper, so from now on we will use the single symbol $\Phi$ to represent both on-shell coherent superstates and off shell superfields. The same is also true for superderivatives $\boldsymbol{D}$ and $\boldsymbol{\overline{D}}$. For reference, we summarize the `replacement rules' in Table \ref{dic}.

\begin{table}[h!]
\begin{center}
\begin{tabular}{ |c|c|c|c| }
 \hline
 \multicolumn{3}{|c|}{Replacement Rule} \\
 \hline
Off-shell & On-shell & Spin-helicity Expression\\
 \hline
$\boldsymbol{\Phi_i} $    & $ \Phi_i $&$\eta_i$   \\
$\boldsymbol{D\Phi_i} $    & $ (Q\Phi)_i $&$\lambda_i$   \\
$\boldsymbol{\overline{D}D\Phi_i} $    & $ (Q^\dag Q\Phi)_i $&$\tilde{\lambda}_i\lambda_i\eta_i$   \\
$\boldsymbol{\Phi^\dag_i} $    & $ \Phi^\dag_i $&$1$   \\
$\boldsymbol{\overline{D}\Phi^\dag_i} $    & $ (Q^\dag\Phi^\dag)_i $&$\tilde{\lambda}_i\eta_i$   \\
$\boldsymbol{D\overline{D}\Phi^\dag_i} $    & $ (QQ^\dag\Phi^\dag)_i $&$\lambda_i\tilde{\lambda}_i$   \\
\hline
\end{tabular}
\end{center}
\caption{This table provides the replacement rule from off-shell superfield to on-shell spin helicity expression under $\eta$-representation. We only list the first several superfields with derivatives. One can easily generalize the result to superfields with any number of derivatives.}
\label{dic}
\end{table}

We should emphasize that we are not defining the group representation of $D$ and $\overline{D}$. Instead, we are identifying the action of superderivatives with differential operators (which are the same as supercharges given the above justification), which allows us to relate off shell fields with on-shell fields. Therefore one need not to worry about the fact that $\{D,Q^\dag\}=0$ (part of the $\mathcal N = 1$ supersymmetry algebra) is not satisfied under this identification.

\section{Hidden U(N) Symmetry and Super Young Diagrams}\label{sec:unsymm}

Having seen that one can express any supersymmetric operator\footnote{in the $\eta$ basis.} in terms of $\lambda,\tilde{\lambda},\eta$, we now show that these variables ($\lambda,\tilde{\lambda},\eta$) transform non-trivially under an internal $U(N)$ symmetry. This symmetry can be used to further restrict the form of the operators and the amplitudes they correspond to.

\subsection{U(N) Symmetry and Amplitudes}\label{unnonsusy}

In a non-supersymmetric theory, we can express the amplitude for $N$ distinguishable, massless particles as $\delta^4(P)f(\lambda_i, \tilde \lambda_i)$, where $i = 1\,...\, N$. Let us strip off the delta function for now and focus on the structure of $f(\lambda_i, \tilde \lambda_i)$. 

The key to the structure of $f(\lambda_i, \tilde \lambda_i)$ is to extend the little group from $U(1)^N$ -- a phase (the little group for a massless superstate) for every particle in the amplitude --  to $U(N)$~\cite{Henning:2019mcv,Henning:2019enq,Li:2020gnx}. Under $U(N)$, the  $\lambda$ transform as the fundamental representation and $\tilde{\lambda}$ transform as anti-fundamental:
\begin{equation}\label{lambda trans}
\lambda\rightarrow u\lambda\ \ \ \ \tilde{\lambda}\rightarrow u^\dag \tilde{\lambda}, \quad u\in U(N).
\end{equation}
This transformation leaves the total momentum unchanged, as one can easily verify:
\begin{equation}
P_{\alpha\dot{\alpha}}=\sum{\lambda_a \tilde{\lambda}_{\dot{\alpha}}}\rightarrow P'_{\alpha\dot{\alpha}}=\sum{(u u^\dag)\lambda_a \tilde{\lambda}_{\dot{\alpha}}}=\sum{\lambda_a \tilde{\lambda}_{\dot{\alpha}}}.
\end{equation}

Under the transformation, the amplitude $f(\lambda_i, \tilde \lambda_i)$ is a tensor product of $U(N)$ fundamentals and anti-fundamentals, which we'll write as
\begin{align}
f(\lambda_i, \tilde \lambda_i) \sim \lambda^{\otimes m} \otimes \tilde{\lambda}^{\otimes n},
\label{eq:fnonsusy}
\end{align}
where $m$ and $n$ are the number of $\lambda$, $\tilde \lambda$.
One well known, diagrammatic way to realize the $U(N)$ tensor product is using Young tableau (YT).  

In principle, the tensor product in Eq.~\eqref{tensor decomposition} contains several Young tableaux. However, we know only certain structures within $f(\lambda_i, \tilde \lambda_i)$ are allowed by kinematic constraints such as symmetrization/antisymmetrization and momentum conservation (IBP). Endowing the $\lambda_i, \tilde \lambda_i$ with $U(N)$ properties, the kinematic constraints manifest in $U(N)$ space. As a result, one can immediately spot the viable terms in $f(\lambda_i, \tilde \lambda_i)$ simply from the shape of the Young diagrams. As one example, spinors $\lambda_i \lambda_j$ must be contracted antisymmetrically in pairs, which in $U(N)$ YT language translates to products of columns that are only two boxes high. Enforcing all the kinematic constraints, the allowed YT have $m/2$ two-box columns (representing the $\lambda $ products) immediately to the right of $n/2$ columns, each of which are $N-2$ boxes high (representing the $\tilde \lambda$ products, expressed in terms of products of $U(N)$ fundamentals using $\epsilon^{12\cdots N}$). An illustrative example is shown below in Fig.~\ref{nonsusy 4}, and we refer the reader to Appendix \ref{Non-Supersymmetric SSYT} and the original work on YT formation of amplitudes, \cite{Henning:2019mcv,Henning:2019enq} for more details. Since its introduction, the (non-supersymmetric) YT technique has been utilized to construct and count operators in a variety of interesting scenarios~\cite{Li:2020xlh,Li:2020tsi,Li:2021tsq,Li:2022tec,Harlander:2023psl,Li:2023wdz,Song:2023jqm,Li:2023cwy}.
\begin{figure}[h!]
\begin{center}
\includegraphics[scale=0.27]{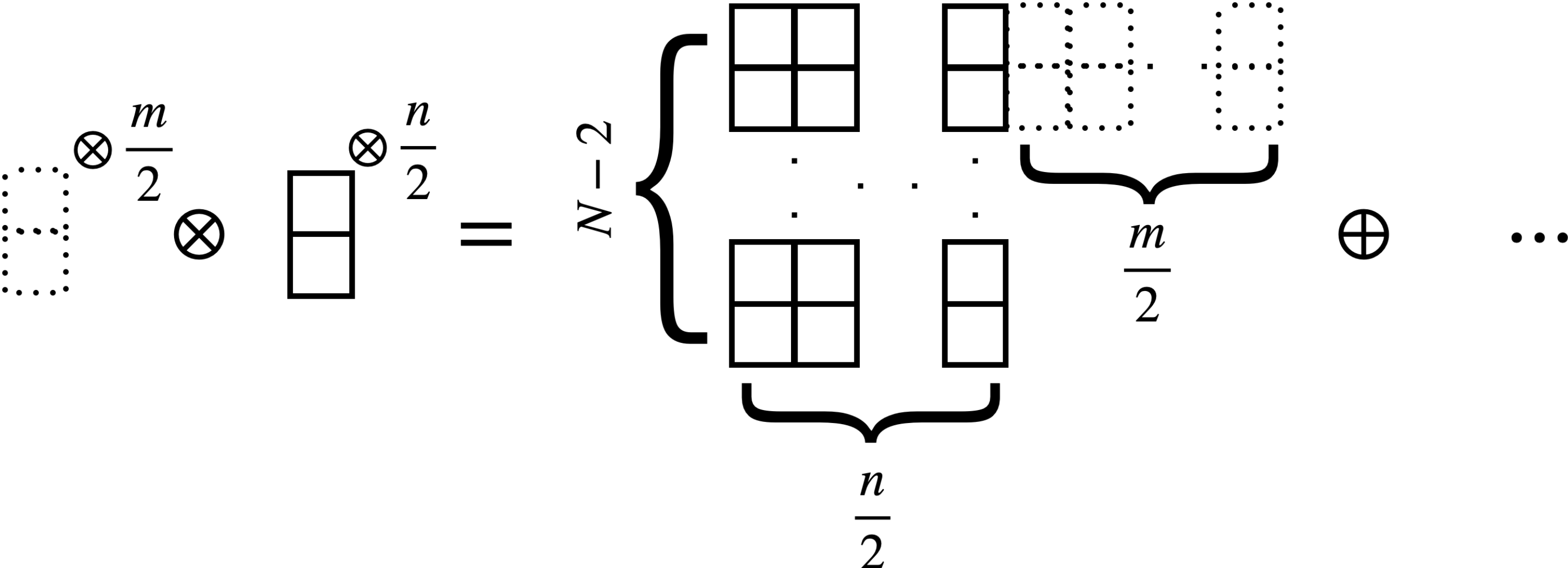}
\end{center}
\caption{Young tableau representation of Eq.~\eqref{eq:fnonsusy}.}
\label{nonsusy 4}
\end{figure}

In the following, we will refer to the structure in Fig.~\ref{nonsusy 4} as `harmonic' YT due to the fact that viable $f(\lambda_i,\tilde{\lambda}_i)$ are harmonics (annihilated by) the  conformal generator $K=\sum_i \frac{\partial}{\partial \lambda_i}\frac{\partial}{\partial \tilde{\lambda_i}}$~\cite{Henning:2019mcv,Henning:2019enq}.

\subsection{U(N) and Superamplitudes}\label{unYT}

To extend the $U(N)$ approach to superamplitudes, we need to know how $\eta$ transforms. We take $\eta$ to transform as a fundamental (and therefore $\partial/\partial \eta$ as an antifundamental) -- the extrapolation of the little group scaling explained in Sec.~\ref{superstate} to $U(N)$ -- as this keeps the supercharges $Q,Q^\dag$ Eq.~\eqref{q rep} invariant. In YT terms, the superamplitude building blocks $\eta, \lambda, \tilde \lambda$ are shown in Fig.~\ref{box rep} below.

\begin{figure}[h!]
\begin{center}
\includegraphics[scale=0.25]{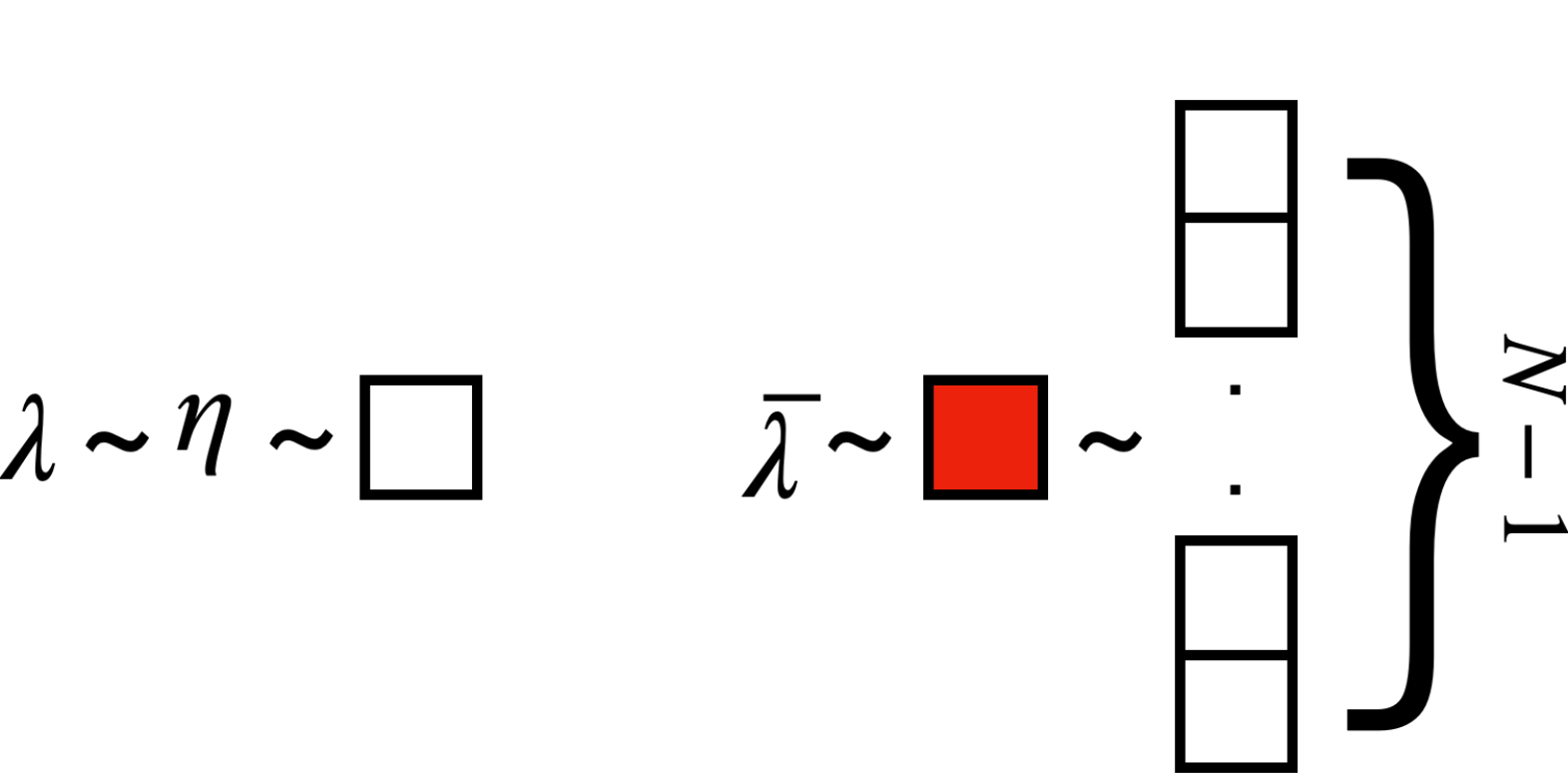}
\end{center}
\caption{$U(N)$ representations for $\lambda, \tilde \lambda$ and $\eta$, where $N$ is the number of superfields in the operator (superstates in the amplitude). The $\lambda, \tilde \lambda$ representations (fundamental, antifundamental) are identical to the non-supersymmetric scenario, and the $\eta$ representation has been chosen to keep the supercharges $Q, Q^\dag$ $U(N)$ invariant. We use a red box to denote antifundamentals, though these can be replaced by fundamentals (white boxes) via $\epsilon^{1,2,...,N}$.}
\label{box rep}
\end{figure}

 We can now use the replacement rules and the $\lambda, \tilde \lambda, \eta$ $U(N)$ properties to express any (higher dimensional) superfield operator $\mathcal{O}$ as a tensor product of $\lambda$s, $\tilde{\lambda}$s and $\eta$s.
\begin{equation}\label{tensor decomposition}
\mathcal{O}\sim \lambda^{\otimes m} \otimes \tilde{\lambda}^{\otimes n} \otimes \eta^{\otimes N_c}.
\end{equation}\label{Nc}
Here, $m,n$ are respectively the numbers of super derivatives $D$ and $\overline{D}$ in $\mathcal{O}$, and 
\begin{equation}
N_c=N_\Phi-m+n,
\end{equation}
where $N_\Phi$ is the number of chiral superfields $\Phi$ in $\mathcal{O}$.\footnote{This is fixed by little group scaling and the details are provided in the companion paper \cite{spinhelicitypaper2}.} For example, consider the following operators with four derivatives and four superfields, two of which are chiral -- ($n = 2, m = 2, N_c = 2)$:
\begin{equation}
\begin{split}
\label{repexample}
&D\Phi_1D\Phi_2\overline{D}\Phi^\dag_3\overline{D}\Phi^\dag_4\sim \lambda_{\alpha1}\lambda_2^\alpha\tilde{\lambda}_{\dot{\alpha}3}\tilde{\lambda}^{\dot{\alpha}}_4\eta_3\eta_4\sim [12] \langle34\rangle\eta_3\eta_4, \\ 
&\overline{D}D\Phi_1D\Phi_2\overline{D}\Phi^\dag_3\Phi^\dag_4\sim \lambda_{\alpha1}\lambda_2^\alpha\tilde{\lambda}_{\dot{\alpha}1}\tilde{\lambda}^{\dot{\alpha}}_3\eta_1\eta_3\sim [12] \langle13\rangle\eta_1\eta_3, \\
&\overline{D}D\Phi_1\overline{D}D\Phi_2\Phi^\dag_3\Phi^\dag_4\sim \lambda_{\alpha1}\lambda_2^\alpha\tilde{\lambda}_{\dot{\alpha}1}\tilde{\lambda}^{\dot{\alpha}}_2\eta_1\eta_2\sim [12] \langle12\rangle\eta_1\eta_2.
\end{split}
\end{equation}

We emphasize that the  tensor product of $U(N)$ representations in Eq.~\eqref{tensor decomposition} is a different beast than the non-supersymmetric $f(\lambda_i, \tilde \lambda_i)$. Equation~\eqref{tensor decomposition} is the tensor product for the superfield operator, which we have associated with the on-shell superstate via the replacement rules in Eqs.~\eqref{field rep} and \eqref{d rep} following the logic explained in Sect.~\ref{On-Shell Versus Off-Shell}.  To form the superamplitude (the analog of the non-supersymmetric $f(\lambda, \tilde \lambda)$), we need to integrate over superspace
\begin{equation}\label{operator action}
\int d^4x \int d^4\theta \mathcal{O}
\sim \int d^4x [(\overline{D}^2D^2 \mathcal{O})|_{\theta=\overline{\theta}=0}],
\end{equation}
where we use the definition of superderivatives $D,\overline{D}$, and total spacetime derivatives vanish upon the $d^4x$ integration. 

While $\mathcal O$ (Eq.~\eqref{tensor decomposition}) itself is not a superamplitude, if we consider its lowest component only -- i.e. ignoring the $\eta$ (or, setting $\theta, \bar\theta \to 0$) -- the resulting component operator is subject to the non-supersymmetric operator $\leftrightarrow$ YT correspondence of Sec.~\ref{unnonsusy}. What this means is that the $\lambda, \tilde\lambda$ part of $\mathcal O$ can be identified as a non-supersymmetric amplitude, and therefore we automatically know its YT form --  harmonic diagrams only! We'll refer to this piece in the following as $g(\lambda, \tilde \lambda)$ (as there is no $\eta$ dependence). 

To turn the formatted supersymmetry operator into a superamplitude, we need to apply the $\bar D^2 D^2$ (see Eq.~\eqref{operator action}). Let's take these one at a time. Given the action defined in Eq.~\eqref{d rep}\footnote{Notice that although the $D^2$ here acts on the entire operator instead of a specific field, its tensor structrue follows Eq.~\eqref{d rep}.}, the result of $D^2$ is to replace two $\eta$ with two $\lambda$ in Eq.~\eqref{tensor decomposition}:
\begin{equation}\label{D2 tensor decomposition}
D^2\mathcal{O}\sim \lambda^{\otimes m} \otimes \tilde{\lambda}^{\otimes n}\otimes (\lambda^{\otimes2}\eta^{\otimes(N_c-2)}) \sim \mathcal O \otimes (\lambda^{\otimes2}\eta^{\otimes(N_c-2)}),
\end{equation}
Importantly, the additional $D^2$ do not affect any of the structure from $\mathcal O$, they just add additional $\lambda$ factors which are contracted among themselves such that the result factors. Said another way,
\begin{align}
\label{step2}
\int d^2\theta\, \mathcal O = g(\lambda, \tilde \lambda)\otimes Z(\lambda, \eta),
\end{align}
where the YT for $g(\lambda, \tilde \lambda)$ has harmonic form and the form of $Z(\lambda, \eta)$ needs to be determined.

To determine the YT form of $Z(\lambda, \eta)$, we use the super-Ward identity Eq.~\eqref{QA}. As $g(\lambda, \tilde \lambda)$ is independent of $\eta$, the product in Eq.~\eqref{step2} is annihilated by $Q$ provided $QZ(\lambda, \eta) = 0$. For $Z \sim \lambda_a \lambda_b \eta_c$ the solution to $QZ = 0$ is known and given by Eq.~\eqref{eq:mabc}. To find the form for more general $Z$, note that in $U(N)$ language, the $m_{abc}$ structure in Eq.~\eqref{eq:mabc} is a totally antisymmetric $U(N)$ object. Extrapolating this property to general $Z$ -- a product of $N_c$ $U(N)$ fundamentals -- one can show that the totally antisymmetric form $Z(\lambda, \eta) = \epsilon^{a_1a_2\cdots  a_{N_c}}\lambda_{a_1}\lambda_{a_2}\eta_{a_3}\cdots\eta_{a_{N_c}}$ is annihilated by $Q$. Explicitly:
\begin{equation}\label{QZ}
\begin{split}
Q Z(\eta,\lambda)
=&\sum_{i=1}^N \lambda_{i} \frac{\partial}{\partial\eta_i}(\epsilon^{a_1a_2\cdots  a_{N_c}}\lambda_{a_1}\lambda_{a_2}\eta_{a_3}\cdots\eta_{a_{N_c}})\\
=&\sum_{i=3}^{N_c}\epsilon^{a_1a_2\cdots  a_{N_c}}\lambda_{a_1}\lambda_{a_2}\lambda_{a_i}\eta_{a_3}\cdots\eta_{a_{i-1}}\eta_{a_{i+1}}\cdots\eta_{a_{N_c}}\\
=& 0
\end{split}
\end{equation}
where we have used the Schouten identity\footnote{Schouten identity is given by $\lambda_i^\alpha\lambda_{j\alpha}\lambda_{k\beta}+\lambda_j^\alpha\lambda_{k\alpha}\lambda_{l\beta}+\lambda_k^\alpha\lambda_{i\alpha}\lambda_{j\beta}=0$ in terms of spin-helicity variables.} in the last line. The $U(N)$ antisymmetric YT form for $Z(\lambda, \eta)$ is represented below in Figure \ref{lambda piece}. 

\begin{figure}[h!]
\begin{center}
\includegraphics[scale=0.3]{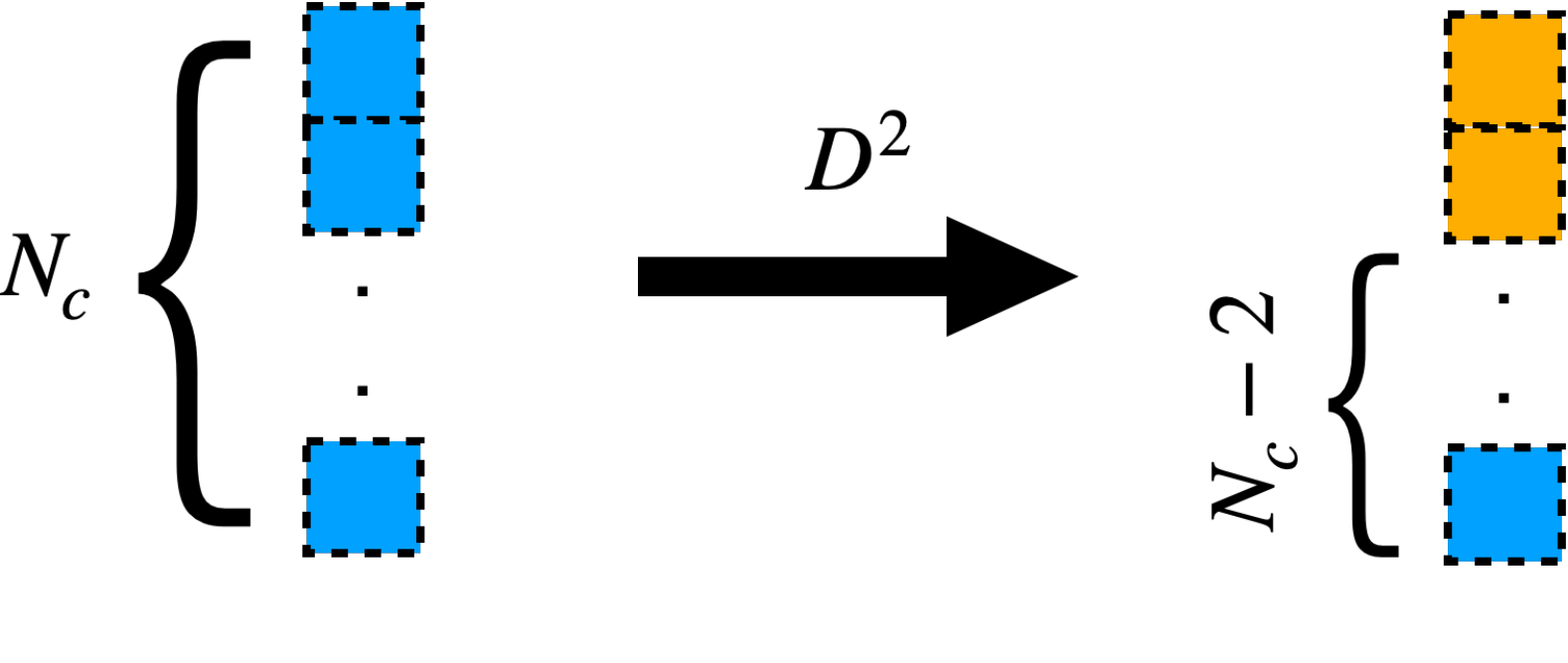}
\end{center}
\caption{The YT form of $Z(\lambda, \eta)$, where we've used blue boxes to represent $\eta$s and orange boxes for $\lambda$s. As $g(\lambda, \tilde \lambda)$ has no $\eta$ dependence, $D^2$ acts solely on the $\eta$ part of the operator, converting two $\eta$ to $\lambda$. In order to satisfy the super-Ward identity $QZ = 0$, the two new $\lambda$ and any remaining $\eta$s must be arranged into the totally antisymmetric combination, a vertical stack of boxes in YT language.}
\label{lambda piece}
\end{figure}

Technically, we have cheated in the last step as $Q$ must annihilate the amplitude, and, as we have stressed, $\int d^2\theta\, \mathcal O$ is not quite an amplitude. The above logic does still apply provided that the last step -- acting with $\bar D^2$ -- reduces to multiplying $\int d^2\theta \mathcal O$ by a factor that is itself annihilated by $Q$. We'll see below that this is the case.

The final step in converting the superfield operator $\mathcal O$ to a superamplitude is to apply the $\bar D^2$, and the remaining property our amplitude must satisfy is the second super-Ward identity, Eq.~\eqref{QdA}. However, from our discussion in Sec.~\ref{sec:superward}, we know that amplitudes in the $\eta$ basis trivially satisfy $Q^\dag A$ by being proportional to $\delta^2(Q^\dag)$. As we have yet to include this factor in this section, the natural suspicion is that the $\delta^2(Q^\dag)$ factor arises as a result of the $\bar D^2$ integral. This turns out to be correct, as can be verified by explicit examples. Including the $\delta^2(Q^\dag)$, we have
\begin{align}
\int d^4\theta\, \mathcal O = \delta^2(Q^\dag)\, g(\lambda, \tilde\lambda)Z(\lambda,\eta).
\end{align}

Diagrammatically, we don't represent the delta function, just as we strip off $\delta^4(P)$ when analyzing non-supersymmetric operators/amplitudes\footnote{The $\delta^4(P)$ factor is also generated for supersymmetric operators upon integrating $d^4x$, and we will omit this factor when considering diagrams.}. This function is associated with (super-)momentum conservation and is the same for all (super-)amplitudes (in the $\eta$ basis). Note that $\delta^2(Q^\dag)$ is also annihilated by $Q$ once we impose momentum conservation, justifying our use of $Q$ acting on $Z(\lambda, \eta)$ as a proxy for $Q$ acting on the full superamplitude.

\begin{figure}[h!]
\begin{center}
\includegraphics[width=0.9\textwidth]{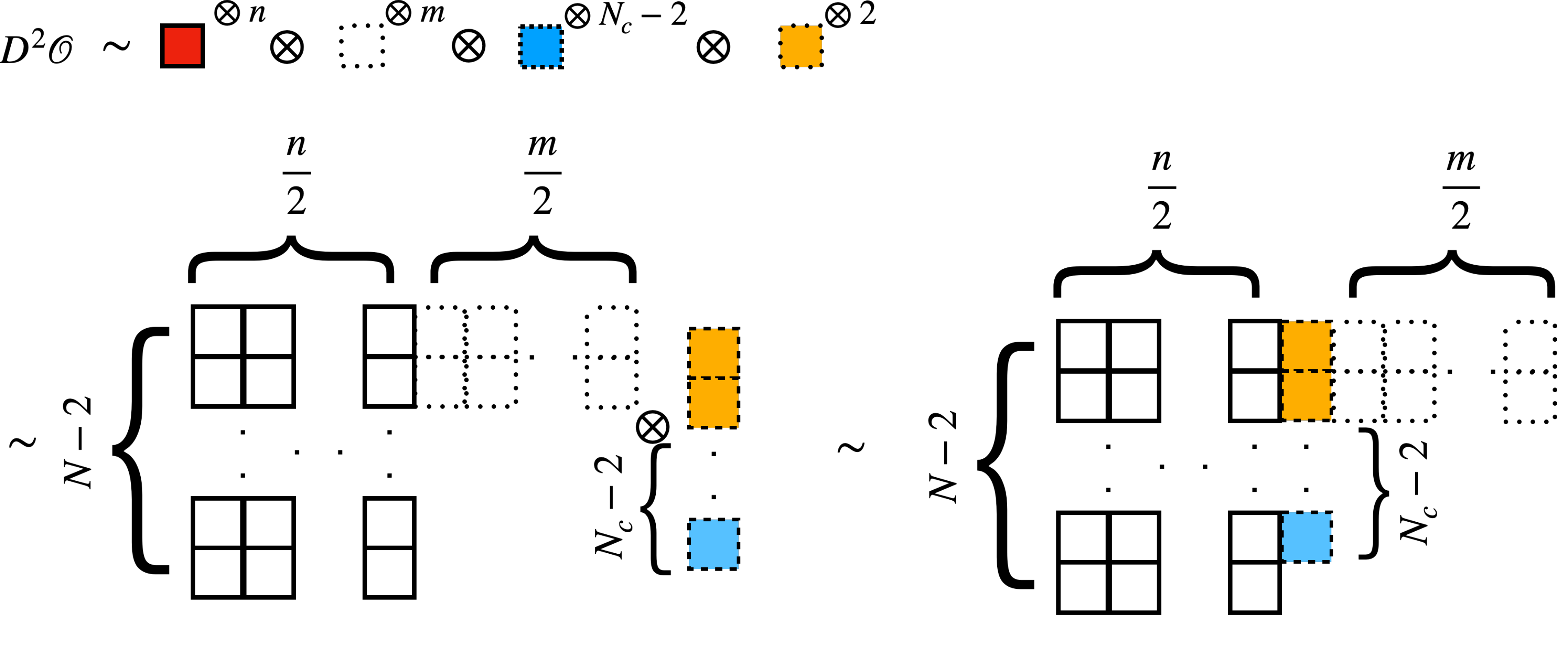}
\end{center}
\caption{Illustration of the Young tableau for a given superfield operator. The ingredients are shown at the top:  $n$ $U(N)$ antifundamentals (red) representing powers of $\tilde \lambda$, $m+ 2$ fundamentals (white and yellow) representing powers of $\lambda$, and $N_c-2$ fundamentals (blue) representing powers of $\eta$. The antifundamentals and $m$ of the fundamentals come entirely from the operator and are arranged in harmonic form, meaning $m/2$ columns of height 2 glued to the right of $n/2$ columns of height $N-2$. To satisfy $Q_\alpha\, A = 0$, the remaining fundamentals (coming from $d^2\theta \sim D^2$) must enter antisymmetrically, meaning a column of height $N_c$. The two pieces (harmonic, operator piece plus antisymmetric $D^2$ piece) are shown separately in the left part of the figure, while in the right figure we  show the only allowed product. }
\label{Master YD}
\end{figure}

 With all $U(N)$ objects accounted for and at last working with a superamplitude, the last step is to combine the $g(\lambda, \tilde \lambda)$ and $Z(\lambda, \eta)$ pieces into a single YT. The only legal form is to slide the $Z(\lambda, \eta)$ column between the $n/2$ antifundamental factors and the $m/2$ fundamental factors, as shown in Fig. \ref{Master YD}. The result is the YT diagram for $\int d^4\theta\, \mathcal O$ without the $\delta^2(Q^\dag)$ factor (i.e. $\int d^2\theta\, \mathcal O$), which is in one-to-one correspondence with the operator itself, and can therefore be used to construct the operator basis, thereby counting and forming all operators with the same field content. We will elaborate upon on exactly how this is done starting from the YT of Fig.~\ref{Master YD} form  in future work~\cite{spinhelicitypaper2}.

 The YT construction we have described in this section holds for any superfield operator provided it contains {\it two or more} chiral superfields\footnote{Technically, we have also assumed two or more antichiral superfields -- as the hermitian conjugate of an operator with zero or one antichiral superfield is an operator with zero or one chiral superfield. For operators with zero or one antichiral fields, the steps proceed as described, except the $Z(\lambda, \eta)$ column has more than $N-2$ boxes and must be placed on the far left of the diagram, rather than in the middle. See Appendix~\ref{eta bar rep}.}. Operators with fewer chiral superfields do not provide enough $\eta$ factors under our replacement rule to survive the $d^2 \theta = D^2$. This issue can be traced back to the fact that we are working in the $\eta$ basis. In addition to governing how superfields are replaced and how the amplitudes behave under the action of $Q$ and $Q^\dag$, this choice also dictated that $d^4\theta \to \bar D^2 D^2$ -- with the $D^2$ acting on the operator first, as opposed to $D^2\bar D^2$. In order to analyze superfield operators with $N_c < 2 $ we need to work in the $\bar \eta$ representation\footnote{Or we can study the amplitude of the hermitian conjugate of our $N_c < 2 $ operator  (which must have $N_c \geq 2$ for all higher dimensional operators we are interested in) then take the hermitian conjugate to restore the original amplitude. See Appendix~\ref{eta bar rep}.}. In the $\bar \eta$ basis for coherent states, antichiral fields are represented by $\bar\eta_i$, and $\overline D$ is $\propto \partial/\partial \bar{\eta}_i$\footnote{More completely, the replacement rules are $\Phi^\dag_i \to \bar\eta, \Phi_i \to 1$, $D_i \to \lambda_i \bar\eta_i$, $\overline D_i \to \tilde \lambda_i \partial/\partial_{\bar \eta_i}$.}. As such, taking $d^4\theta = D^2\overline D^2$, the action of $\overline D^2$ is now non-zero on any operator with two or more antichiral fields, regardless of the number (including zero or one) of chiral superfields. In other words, the logical flow is identical to the construction here with the roles of chiral/antichiral superfields and $Q/Q^\dag$ (and therefore $D/\overline D$), swapped. We show an example calculation using the $\bar \eta$ state basis in Appendix~\ref{eta bar rep}.\footnote{It would seem that the scenario of a higher dimensional operator consisting of only one chiral, one antichiral, and several $D/\overline D$ causes problems as neither the $\eta$ nor $\bar \eta$ bases will work. However, all such $D^m\overline D^n \Phi \Phi^\dag$ terms vanish upon IBP and usage of the EOM.}

\subsubsection{Examples}
In this section, we go through a real example, expanding out the action of all $D, \bar D$ to verify the form in Fig.~\ref{Master YD}. Consider the dimension seven (before integrating over $d^4 \theta$) operator $D\Phi_1D\Phi_2\Phi_3\overline{D}\Phi^\dag_4\overline{D}\Phi^\dag_5$, which has $N = 5, n = m = 2, N_c = 3$. Using our replacement rules, this operator becomes:
\begin{equation}
 [12] \langle45\rangle\eta_3\eta_4\eta_5.
\end{equation}
Next we act with the $D^2$. Let's do things one $D$ one at a time:
\begin{equation}
\begin{split}
&D_\beta(D_\alpha\Phi_1D^\alpha\Phi_2\Phi_3\overline{D}_{\dot{\alpha}}\Phi^\dag_4\overline{D}^{\dot{\alpha}}\Phi^\dag_5)\\
\sim&D\Phi_1D\Phi_2D_\beta\Phi_3\overline{D}\Phi^\dag_4\overline{D}\Phi^\dag_5+D\Phi_1D\Phi_2\Phi_3D_\beta\overline{D}\Phi^\dag_4\overline{D}\Phi^\dag_5+D\Phi_1D\Phi_2\Phi_3\overline{D}\Phi^\dag_4D_\beta\overline{D}\Phi^\dag_5,
\end{split}
\end{equation}
where we make use of the definition of anti-chiral superfield $D\Phi^\dag=0$, and terms that contain $D^2\Phi$ are removed because of EOM redundancies. Notice that all contracted indices are hidden on the second line.  Acting with the second $D$ gives \footnote{Notice that an additional minus sign appears when moving $\eta$ pass $\lambda$ and it cancels the minus sign in front of the first term on second line.}:
\begin{align}
&D^\beta(D\Phi_1D\Phi_2D_\beta\Phi_3\overline{D}\Phi^\dag_4\overline{D}\Phi^\dag_5+D\Phi_1D\Phi_2\Phi_3D_\beta\overline{D}\Phi^\dag_4\overline{D}\Phi^\dag_5-D\Phi_1D\Phi_2\Phi_3\overline{D}\Phi^\dag_4D_\beta\overline{D}\Phi^\dag_5) \nonumber \\
\sim&-D\Phi_1D\Phi_2D_\beta\Phi_3D^\beta\overline{D}\Phi^\dag_4\overline{D}\Phi^\dag_5+D\Phi_1D\Phi_2D_\beta\Phi_3\overline{D}\Phi^\dag_4D^\beta\overline{D}\Phi^\dag_5-D\Phi_1D\Phi_2D_\beta\Phi_3D^\beta\overline{D}\Phi^\dag_4\overline{D}\Phi^\dag_5 \nonumber \\
+&D\Phi_1D\Phi_2\Phi_3D_\beta\overline{D}\Phi^\dag_4D^\beta\overline{D}\Phi^\dag_5+D\Phi_1D\Phi_2D_\beta\Phi_3\overline{D}\Phi^\dag_4D^\beta\overline{D}\Phi^\dag_5+D\Phi_1D\Phi_2\Phi_3D_\beta\overline{D}\Phi^\dag_4D^\beta\overline{D}\Phi^\dag_5 \nonumber \\
& \propto\,  [12][34] \langle45\rangle\eta_5+[12][53] \langle45\rangle\eta_4+[12][45] \langle45\rangle\eta_3 \nonumber \\
& =\, [12] \langle45\rangle([34]\eta_5+[45]\eta_3+[53]\eta_4),
\end{align}
where in the last step we transform the expression into spin-helicity form (via replacement rules) and omit any overall constants. Clearly, $[12]\langle45\rangle$ corresponds to the left part in Figure \ref{Master YD} while the terms in parentheses is Figure \ref{lambda piece}. This brings us to the last step --  acting $\overline{D}^2$ on the whole expression:
\begin{equation}
\overline{D}^2(-D\Phi_1D\Phi_2D\Phi_3D\overline{D}\Phi^\dag_4\overline{D}\Phi^\dag_5+D\Phi_1D\Phi_2D\Phi_3\overline{D}\Phi^\dag_4D\overline{D}\Phi^\dag_5
+D\Phi_1D\Phi_2\Phi_3D\overline{D}\Phi^\dag_4D\overline{D}\Phi^\dag_5).
\label{eq:lastD2}
\end{equation}
Focusing on the first term for brevity, expanding out the $\overline D^2$ and converting to spinor helicity form using the replacement rules, we find (again up to an overall constant prefactor):
\begin{align}
\overline{D}^2(-D\Phi_1D\Phi_2D\Phi_3D\overline{D}\Phi^\dag_4\overline{D}\Phi^\dag_5)=&\,[12]\langle 4 5 \rangle [34]\eta_5 (\langle 1 2 \rangle \eta_1 \eta_2 + \langle 1 3 \rangle \eta_1 \eta_3 + \langle 1 4 \rangle \eta_1 \eta_4 \nonumber \\
& + \langle 2 3 \rangle \eta_2 \eta_3 + \langle 2 4 \rangle \eta_2 \eta_4 + \langle 3 4 \rangle \eta_3 \eta_4 ) \nonumber \\
& = [12]\langle 4 5 \rangle [34]\eta_5\, \Big( \sum_{i < j}^{5}\langle ij\rangle\eta_i\eta_j \Big)
\end{align}
where we've used the fact that $\eta^2_5 = 0$ to convert the terms in parenthesis in the first line into the sum on the last line, which we recognize as $\delta^2(Q^\dag)$. The same manipulation can be performed on the other terms in Eq.~\eqref{eq:lastD2}, leading to the full result:
\begin{equation}
\label{exampleend}
\begin{split}
D\Phi_1D\Phi_2\Phi_3\overline{D}\Phi^\dag_4\overline{D}\Phi^\dag_5\, \to\, &[12] \langle45\rangle([34]\eta_5+[45]\eta_3+[53]\eta_4)\Big(\sum_{i<j}^{5}\langle ij\rangle\eta_i\eta_j\Big)\\
\to\, &[12] \langle45\rangle D^2(\eta_3\eta_4\eta_5)\delta^2(Q^\dag),
\end{split}
\end{equation}
where in the last line we've written things in a way that clearly factors the $g(\lambda, \tilde \lambda), Z(\lambda, \eta)$ pieces.

Translating Eq.~\eqref{exampleend} into a YT,  the $\langle \rangle$ and $[]$ products become columns of height 2, though we re-express the product of antifundamentals as the antisymmetric product of fundamentals using the $U(N)$ $\epsilon$ symbol. These components are arranged in harmonic form, then combined with a three-box column from $D^2(\eta_3\eta_4\eta_5)$ as shown below in Fig.~\ref{YTexamplefull}\footnote{Fig.~\ref{YTexamplefull}, strickly speaking, as a harmonic represents the operator up to the momentum conservation. See \cite{Henning:2019mcv,Henning:2019enq} for details on harmonics.}. 
\begin{figure}[h!]
\centering
\includegraphics[width=0.75\textwidth]{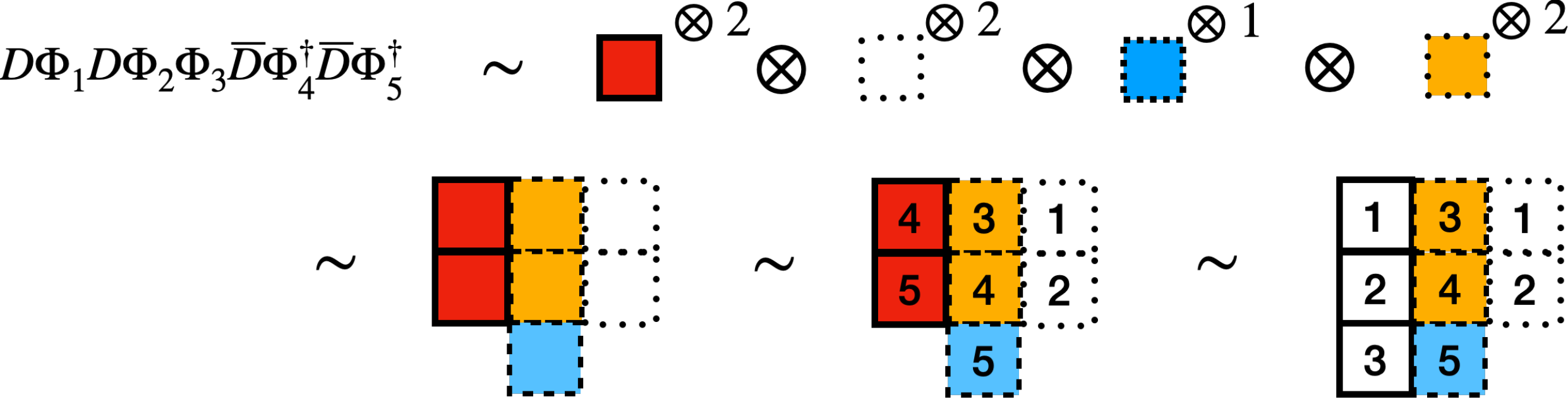}
\caption{Young tableau for the operator $D\Phi_1D\Phi_2\Phi_3\overline{D}\Phi^\dag_4\overline{D}\Phi^\dag_5$, where the building blocks are colored according to Fig.~\ref{Master YD}. As there are two $D, \overline D$, the left and right columns are two boxes high, though the left column becomes three boxes high when we convert $U(N)$ antifundamantals to fundamentals (red boxes to white). The middle column is three boxes high and set by the number of chiral superfields. The entries of the left and right columns are straightforward and follow which fields the $D, \overline D$ act upon. The entries in the middle column can be understood either from the operator's spinor-helicity form, or by recalling that the middle column arises from $d^2\theta = D^2$ and  EOM eliminate $D^2\Phi_1, D^2\Phi_2$.}
\label{YTexamplefull}
\end{figure}
This shape precisely matches what the procedure described in Sec.~\ref{unYT} prescribes for $N = 5, n = m = 2, N_c = 3$.

Specifying the exact operator, rather than just the $N,n,m,N_c$ values does get us more information --  namely we can fill in the boxes with numbers indicating which field/spinor product they represent. For this example, the rightmost column is filled with $1,2$ as it corresponds to $[12]$, and the middle column, corresponding to $Z(\lambda, \eta)$, or $D^2$ in the notation of Eq.~\eqref{exampleend} is filled with $3,4,5$. The fields in the antifundamental are $4,5$, though to convert antifundamental indices to fundamental we apply $\epsilon^{12345}$, getting us $1,2,3$. The final, filled in YT is shown in the rightmost part of Fig.~\ref{YTexamplefull}. Two other operator $\leftrightarrow$ YT example translations are shown below in Fig.~\ref{Example 1}. 

Translating from the YT back to the operator -- the inverse of what we showed in the example above -- is also straightforward. The most direct approach is to first pick out the middle column, which has height $N_c$. From $N_c$, the number of columns that surround it ($n, m$), and the height of the leftmost column ($N-2$), we can work out the number of chiral and antichiral fields. Next, stripping away the middle column, we have the harmonic diagram that corresponds to the operator alone. Identifying entries in the righthand columns of the harmonic diagram with $D$ and entries in the lefthand columns as $\overline{D}$, we can write the operator. 
\begin{figure}[h!]
\begin{center}
\includegraphics[scale=0.4]{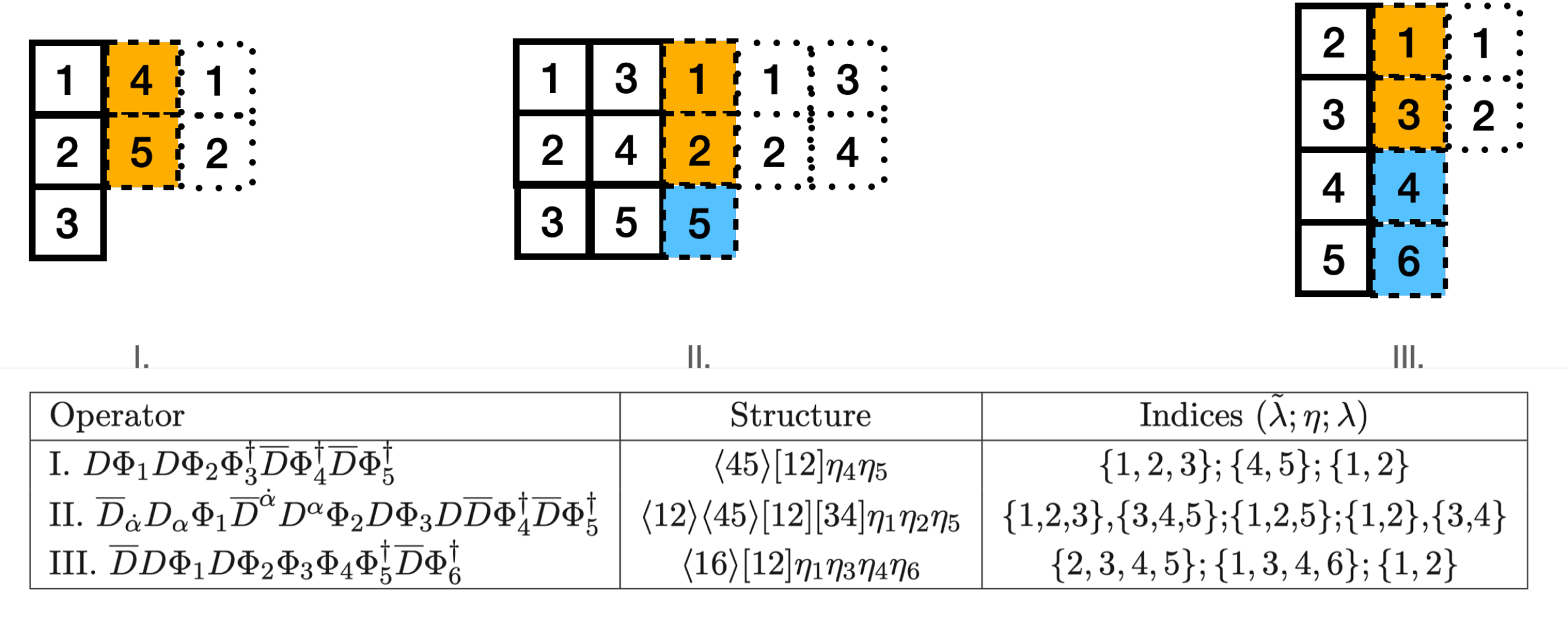}
\end{center}
\caption{Three further examples showing the operator $\leftrightarrow$ YT translation. The YT are shown above a table which contains each of the operators in terms of superfields, the operators' spinor helicity form (structure column), and the grouping of the particle number indices in the YT columns (indices column). }
\label{Example 1}
\end{figure}
 Picking the rightmost diagram as an example of working from YT to operator,  we see $N = 6$ and $N_c = N_\Phi =4$, as $m = n = 2$. The rightmost column contains 1,2, which we translate as $D$ acting on $\Phi_1, \Phi_2$. The leftmost column, once reverted to anti-fundamentals, contains 1 and 6 implying $\overline D$ on $\Phi_1, \Phi_6$. Putting this information together, we arrive at the operator shown. 
 
 As a final remark, we emphasize that all operators with the same $N,n,m,N_c$ values will have the same YT shape and will only differ in how the boxes are filled in. Enumerating the allowed, independent ways to fill in the YT is identical to finding the basis of operators for  a given $N,n,m,N_c$. For non-supersymmetric theories, this counting is accomplished by putting the YT in SSYT form~\cite{Li:2020gnx,Henning:2019mcv,Henning:2019enq}. In a follow up publication, we will show how to do the enumeration for the supersymmetric case.

\subsubsection{Summary of Tableau Construction Rules}\label{rules and examples}
Having worked through some examples, we summarize our procedure to go from superfield operators containing two or more chiral (and antichiral) superfields to a filled-in YT.

\begin{itemize}
\item[1.)] First, identify the structure of $\lambda,\tilde{\lambda},\eta$ of the original superfield operator, i.e. Eq. \eqref{tensor decomposition} using the replacement rules in Table~\ref{dic}.
\item[2.)] Spinor products of $\lambda_i$ become two-box columns forming the rightmost piece of the YT. These boxes should be filled with indices indicting which $D$ are contracted, e.g.  boxes with entries 1 and 2 for $D_\alpha\Phi_1 D^\alpha \Phi_2$. 
\item[3.)] The $\eta_i$ become a column of height $N_c$ (where $N_c \ge 2)$. This column goes immediately to the left of the boxes in Step 2.), and the boxes are filled with the $\eta_i$ indices. This column arises from the application of $D^2$, and while $D^2$ replaces two $\eta_i$ with $\lambda_i$, this does not affect the shape or entries of the YT in any way. \item[4.)] To the left of Step 3.), add boxes for the spinor products of $\tilde \lambda_i$. These products always involve two $\tilde \lambda$ and therefore two boxes, however to convert anti-fundamental boxes to fundamental boxes, these products become columns of height $N-2$, where $N$ is the total number of fields in the operator. For the same reason, the numbers entered into the box are the complement from $\{1,2,\cdots N\}$, e.g. $\{1,2,3\}$ for an operator with $N=5$ containing $\langle 45 \rangle$.
\end{itemize}

The `middle' column, which corresponds to $Z(\lambda, \eta)$ in the notation of Sec.~\ref{unYT} is the main difference between the supersymmetric YT and YT in non-supersymmetric theories. Its role is to make entire operator/amplitude supersymmetric by enforcing the Ward identity $Q\,A = 0$.

\section{Conclusion and Discussion}
 In this paper we have shown how the Young tableau method of constructing non-factorizable, non-supersymmetric amplitudes -- which are in one-to-one correspondence with higher dimensional operators -- can be extended to the case of superamplitudes/operators in $\mathcal{N}=1$ supersymmetry with massless, distinguishable fields. In particular, the same $U(N)$ symmetry (where $N$ is the number of fields in the operator/states in the amplitude) that helps organize non-supersymmetric amplitudes is present in the supersymmetry case. 

After reviewing the coherent state picture for superstates, we developed a replacement rule which allows one to translate off-shell superfield operators into products of spinor helicity and Grassmann coherent state ($\eta$) variables. These rules are akin to the non-supersymmetric replacements in Eq.~\eqref{eq:replaceonshell}, though a bit more subtle as supermultiplets contain several different fields. Endowing the spinor/Grassmann $\eta$ variables with $U(N)$ transformation properties, operators become tensor products of $U(N)$ fundamentals and antifundamentals and can be arranged diagrammatically using YT. Products of $\lambda_i$ and $\tilde \lambda_i$ are treated exactly as in the non-supersymmetric case, meaning their YT are restricted to tableau with harmonic form. The Grassmann variables $\eta_i$ form an additional column in the YT, which resides between the $\tilde \lambda_i$ products and the $\lambda_i$ products. This extra piece (along with an overall factor of $\delta^2(Q^\dag)$) enforces the supersymmetric Ward identities and arises from the $d^4\theta$ integration that converts a superfield operator into an amplitude. Finally, to complete the translation from higher dimensional superfield operator to YT, the boxes of the tableau can be filled in with the labels of the spinor/Grassmann variables involved, e.g. boxes corresponding to $\lambda_1 \lambda_2$ are filled with $1$ and $2$, etc. A flowchart summarizing these steps is shown below in Fig. \ref{flow}. 
\begin{figure}[h!]
\begin{center}
\includegraphics[scale=0.28]{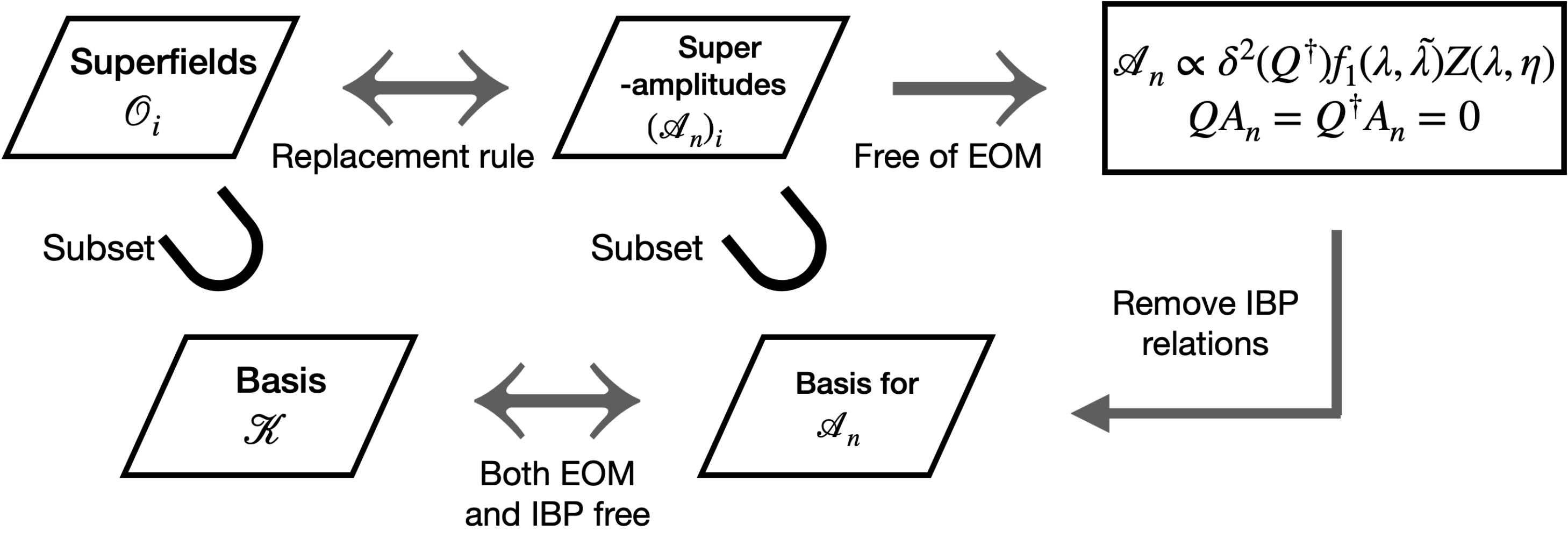}
\end{center}
\caption{Given a set of superfields $\{\mathcal{O}\}$ we can form the corresponding superamplitudes $\{\mathcal{A}\}$ in terms of $\lambda,\tilde{\lambda},\eta$. Relating spin-helicity variables with fundamental representations of the hidden $U(N)$ group allows a Young tableaux construction. Only one unique Young diagram survives when considering constraints from super-Ward identities. While most of the examples in the text involve chiral/antichiral superfields, the method can be applied to operators including (massless, distinguishable) vector superfields as well (see Appendix~\ref{app:vector}).}
\label{flow}
\end{figure}

The steps depicted in Fig.~\ref{flow} assume that the superfield operator contains at least two chiral superfields. This caveat can be traced back to the fact that there are two different, though equivalent coherent state formulations. Figure~\ref{flow}, and the bulk of this paper, assumes a convention where superstates begin with the component field with the highest helicity (the $\eta$ basis). The basis choice affects how $Q, Q^\dag$, and therefore $D, \overline D$, act on the states. For operators with zero or one chiral superfield, we must switch conventions to the $\bar\eta$ basis, where coherent states begin with the lowest helicity. In the $\bar\eta$ basis, Fig.~\ref{flow} still applies, though with the roles of $Q$ and $Q^\dag$ swapped. An explicit example using the $\bar \eta$ basis is shown in Appendix ~\ref{eta bar rep}.

For any particular higher dimensional superfield operator, the  super Young tableau is uniquely determined, so it's natural to ask if we can find a basis for all super Young diagrams which have the same shape (same $N, m, n, N_c$ values, in the language of Sec.~\ref{unYT}) but different fillings of the boxes. For non-supersymmetric theories, we can find a special basis by filling in numbers following the so-called semi-standard Young tableau (SSYT) format. We will show a similar SSYT approach for constructing an operator basis for both distinguishable and indistinguishable superfields in a forthcoming paper \cite{spinhelicitypaper2}.

\acknowledgments

This is partially supported by the National Science Foundation under Grant Number PHY-2112540.

\appendix

\section{Non-Supersymmetric SSYT}\label{Non-Supersymmetric SSYT}
In this appendix we review the Young tableaux approach to constructing non-factorizable amplitudes/higher dimensional operators in non-supersymmetric theories.  For more depth, we refer interested readers to  \cite{Henning:2019enq,Henning:2019mcv} .

In non-supersymmetric case, non-factorizable amplitudes $A$ formed from distinguishable, massless fields are a scalar functions of $\lambda,\tilde{\lambda}$:
\begin{equation}
A\propto f(\lambda,\tilde{\lambda}),
\end{equation}
which one can express as a tensor product following Eq.~\eqref{lambda trans}:
\begin{equation}\label{non-susy amplitude tensor}
A\sim \lambda^{\otimes m} \otimes \tilde{\lambda}^{\otimes n}.
\end{equation}
Here, $m,n$ represent the total number of spinorial indices $\alpha,\dot{\alpha}$ (before contraction) the amplitude contains. The building blocks in this tensor product are spinor helicity pairs: $\epsilon^{\alpha_1\alpha_2}\lambda^i_{\alpha_1}\lambda^j_{\alpha_2} = [ij],\ \epsilon^{\dot{\alpha}_1\dot{\alpha}_2}\tilde{\lambda}_{i,\dot{\alpha}_1}\tilde{\lambda}_{j,\dot{\alpha}_2} = \langle ij \rangle $, where the epsilon symbol is the usual $2 \times 2$ antisymmetric tensor and $i,j$ label which particle we're referring to and run from  $1$ to $N$ ($N$ being the number of particles in the amplitude).

Next, interpret the $i,j$ as indices of $U(N)$ multiplets rather than just particle labels -- thereby taking $\lambda, \tilde \lambda $ as objects that transform under $U(N)$ as well as Lorentz symmetry. Specifically, we take  upper $i,j,\cdots$ as fundamental indices ($\lambda$ as a $U(N)$ fundamental) and lower $i,j,\cdots$ as antifundamental indices ($\tilde \lambda$ as an antifundamental). 

\begin{figure}[h!]
\begin{center}
\includegraphics[scale=0.35]{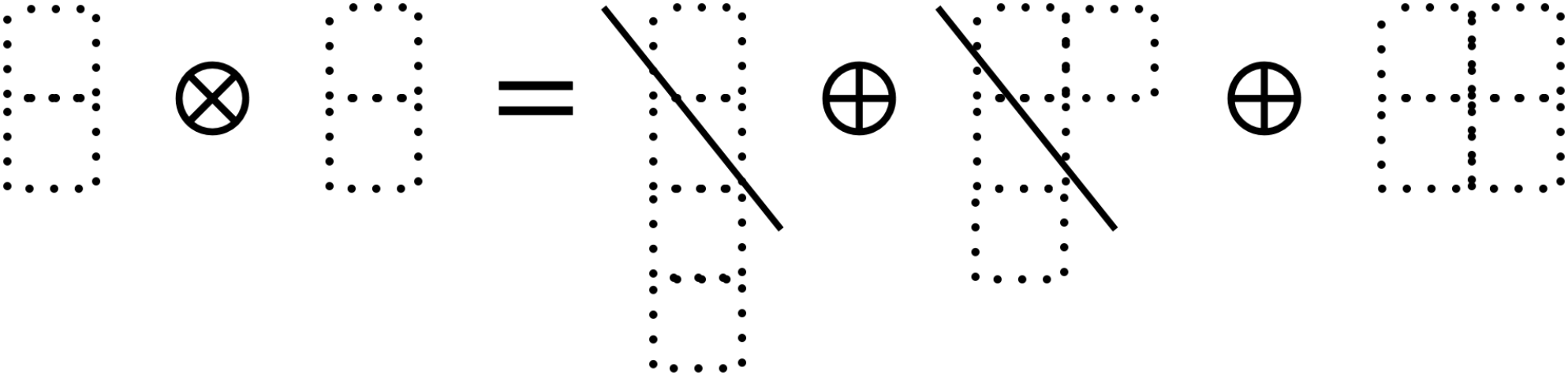}
\end{center}
\caption{The first two Young tableaux on the RHS vanish because of the antisymmetric property of $\lambda$.}
\label{nonsusy 2}
\end{figure}

\begin{figure}[h!]
\begin{center}
\includegraphics[scale=0.25]{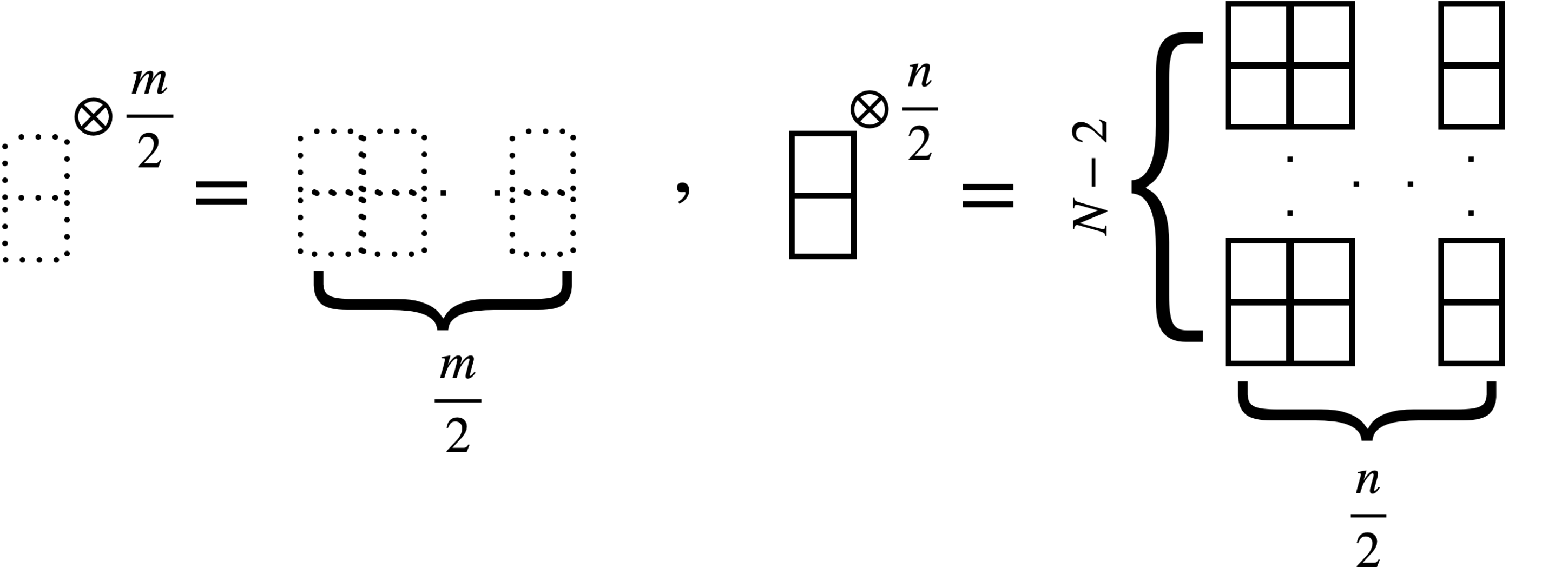}
\end{center}
\caption{Box representations of $\lambda^{\otimes m}$ and $\tilde{\lambda}^{\otimes n}$ pieces separately, where $m,n$ are the total number of each spin-helicity variable, while $N$ is the number of distinguishable fields.}
\label{nonsusy 3}
\end{figure}

Having equipped $\lambda, \tilde\lambda$ with $U(N)$ representations,  we can diagrammatically study the tensor product Eq.~\eqref{non-susy amplitude tensor} using Young tableaux. The basic building block $[ij]$ becomes a column of height two. Taking the product of two `two-boxes' following the usual manipulation of Young tableaux, we find Fig. \ref{nonsusy 2}.
Among the three Young diagrams in the tensor product of two building blocks, only the last one survives because of the antisymmetric nature of spin-helicity variables. This continues for products of more $[ij][$ (or, separately for $\langle ij \rangle$) so that the net $\lambda^{\otimes m}, \tilde \lambda^{\otimes n}$ pieces of the Eq.~\eqref{lambda trans} are shown in Fig.~\ref{nonsusy 3}. Note that in Fig.~\ref{nonsusy 3} we go from anti-fundamental representation to the fundamental representation in $U(N)$ group by contracting the Levi-Civita symbol, e.g $\langle i j \rangle \to \langle i j \rangle\epsilon^{ijk_1 \cdots k_{N-2}} = \tilde\lambda_{i, \dot\alpha} \tilde \lambda^{\dot\alpha}_j \epsilon^{ijk_1 \cdots k_{N-2}}$ (with $i,j$ summed over).

The remaining step is to take the tensor product of the  $\lambda^{\otimes m}$ and $\tilde{\lambda}^{\otimes n}$ pieces. The result, as we can imagine, contains many tableaux.  However, many of the tableaux correspond to operators that are related to other operators with the same field content (and therefore not independent) via IBP. With the YT organization, these redundant terms are easily spotted. Specifically, any YT with additional boxes under the $n/2$ columns of height $N-2$ (thereby making a column of height $N-1$) can be show to be proportional to the total momentum, indicating that the operator corresponding to the diagram is a total derivative/redundant by IBP. Throwing out these pieces, the only remaining tableau in the tensor product contains  $m/2$ columns of height two stacked to the right of the $n/2$ columns of height $N-2$, as shown in Fig. \ref{nonsusy 4}. A more formal way to state which YT are allowed is that viable YT correspond to amplitudes $A$ that are harmonics of the conformal generator $K_{\alpha{\dot{\alpha}}}=\sum_{i=1}^N \frac{\partial}{\partial\lambda_{i\alpha}}\frac{\partial}{\partial\tilde{\lambda}_{i\dot{\alpha}}}$,
\begin{equation}
K_{\alpha{\dot{\alpha}}}A=0.
\end{equation}
See \cite{Henning:2019mcv,Henning:2019enq} for the proof of this statement.

\section{Operators with zero or one chiral superfield}\label{eta bar rep}

If the higher dimensional superfield operator of interest contains zero or one chiral superfield, a direct application of the translation to YT (as summarized in Sec.~\ref{rules and examples}) does not work. Operators of this type have fewer than two $\eta$, and are therefore annihilated when we apply $d^2\theta = D^2$. This apparent annihilation an artifact of working in the $\eta$ basis, so to study operator with $< 2$ chiral superfields we need to switch to the $\bar\eta$ basis.

Let us look at an example $D\Phi_1D\overline{D}\Phi_2^\dag \overline{D}\Phi_3^\dag\Phi_4^\dag$, which has only one chiral superfield. Following the $\eta$ basis translation rule, this operator becomes $\langle23\rangle[12]\eta_3$, which, as expected, doesn't have enough $\eta$ powers to survive $D^2$. Examining the component field expansion, the operator is clearly not equal to zero:
\begin{equation}
\begin{split}
\int d^4\theta D\Phi_1D\overline{D}\Phi_2^\dag \overline{D}\Phi_3^\dag\Phi_4^\dag \supset \partial\psi_1\partial^2\phi^*_2\psi^\dag_3\phi_4^*-\partial\phi_1\partial^2\phi^*_2\partial\phi^*_3\phi_4^*+\cdots
\end{split}
\end{equation}
where we've dropped all indices ($\partial^2$ here is short for $\partial_{\{\mu,\nu\}}$) and only written the first few terms.

To align the spinor helicity/YT result with the component field result, we need to work in the $\bar\eta$ basis for coherent states. In the $\bar\eta$ basis, states are built up from their lowest helicity components, generating superwavefunctions
\begin{align}
\Phi_i &= \phi_i + \bar\eta_i \psi_i \nonumber \\
\Phi^\dag_i &= \psi^\dag_i + \bar\eta \phi_i^*
\end{align}
This inspires a superfield $\to$ on-shell replacement rule for $\Phi, \Phi^\dag, D, \overline D$:
\begin{align}
\Phi_i & \to 1 \nonumber \\
\Phi^\dag & \to \bar\eta_i \nonumber \\
D_i &\to \lambda_i \bar\eta_i \nonumber \\
\overline D_i &\to \tilde\lambda_i \partial/\partial\bar\eta_i, \nonumber 
\end{align}
so that $D\Phi_i \to \lambda_i \bar\eta_i$, $\overline D \Phi^\dag_i \to \tilde\lambda_i$ and the usual chiral/antichiral conditions and EOM are satisfied. Comparing these rules with Table~\ref{dic}, we see  the spinor-helicity pieces match, but the Grassmann variables have moved. Under these rules, our example operator translates to
\begin{align}
\label{etabar}
D\Phi_1D\overline{D}\Phi_2^\dag \overline{D}\Phi_3^\dag\Phi_4^\dag \to [12]\langle 2 3 \rangle \bar\eta_1 \bar\eta_2 \bar\eta_4 
\end{align}
To convert this operator to a superamplitude, we apply $d^4\theta = D^2\overline D^2$. In it's $\bar \eta$ form, Eq.~\eqref{etabar} has enough $\bar\eta$ powers to survive $\overline D^2 \propto \partial/\partial\bar\eta_i\partial/\partial\bar\eta_j$, while the $D^2$ will lead to an overall factor of $\delta^2(Q)$ -- just as we saw in Sec.~\ref{unYT} but with the roles of $D/\overline D$ reversed. The net result is
\begin{align}
\label{eq:orig}
\int d^4\theta D\Phi_1D\overline{D}\Phi_2^\dag \overline{D}\Phi_3^\dag\Phi_4^\dag \to  \delta^2(Q)[12]\langle 2 3 \rangle (\langle 1 2\rangle \bar\eta_4 + \langle 2 4 \rangle \bar\eta_1 + \langle 4 1\rangle \bar\eta_2)
\end{align}
with a shorthand for the last term is $\overline D^2(\bar\eta_1\bar\eta_2 \bar\eta_4)$.

Of course, if a higher dimensional operator $\mathcal O$ has fewer than two chiral superfields, then its hermitian conjugate $\mathcal O^\dag$ will automatically contain more than two chiral superfields and can be translated using the $\eta$ basis. Carrying this out for our example, $D\Phi_1D\overline{D}\Phi_2^\dag \overline{D}\Phi_3^\dag\Phi_4^\dag \xrightarrow[h.c]{} \overline D\Phi^\dag_1\,\overline DD\Phi_2\, D\Phi_3\,\Phi_4$, and
\begin{align}
\label{eq:thehc}
\int d^4\theta D\Phi^\dag_1\,\overline DD\Phi_2\, D\Phi_3\,\Phi_4 \to \delta^2(Q^\dag)\langle 12 \rangle [23] ([12]\eta_4 + [24]\eta_1 + [41]\eta_2)
\end{align}
Comparing Eq.~\eqref{eq:orig} and \eqref{eq:thehc} with each other, we see that they are herimtian conjugates provided we send $\eta_i \leftrightarrow \bar\eta_i$. 

The YT for $\overline D\Phi^\dag_1\,\overline DD\Phi_2\, D\Phi_3\,\Phi_4$ (and therefore for $D\Phi_1D\overline{D}\Phi_2^\dag \overline{D}\Phi_3^\dag\Phi_4^\dag$) is a little unusual in that the $D$ and $\overline D$ columns each have height two, but, as there are three chiral superfields, the middle column has height three. To make a legal YT from these pieces, the column corresponding to $d^2\theta$ (usually the middle) needs to be placed on the far left. This unconventional ordering only occurs when the operator has fewer than two antichiral fields  (meaning that its hermitian conjugate has fewer than two chiral superfields).

\section{Example with vector superfields}\label{app:vector}

In this appendix we give an example of the YT technique for operators containing vector superfields. Following the logic that lead to Eq.~\eqref{eq:swavefn}, we can determine the superwavefunctions for vector superfields in the $\eta$ basis:

\begin{align}
W = G_+ + \eta\, \chi+ \nonumber \\
\overline W = \chi_- + \eta\, G_-
\end{align}
where $G_\pm$ are the $\pm$ helicity spin-1 gauge fields and $\chi_\pm$ are the helicity $\pm1/2$ gauginos (and we have suppressed spinor indices). This organization implies the following replacement for $W, \overline W, DW$ and $\overline D \overline W$ in on-shell superamplitudes:
\begin{align}
W_{i,\alpha} &\to \eta_i \lambda_{i,\alpha}\, ,\quad D_{i,\beta}W_{i,\alpha} \to (\lambda_{i,\beta}\frac{\partial}{\partial \eta_i})( \eta_i \lambda_{i,\alpha}) \to \lambda_{i,\beta}\lambda_{i,\alpha}\, \quad \overline D^{\dot\alpha}_i W_{i,\alpha} \to \tilde\lambda^{\dot\alpha}_i \lambda_{i,\alpha}\eta^2_i = 0 \nonumber \\
\overline{W}^{\dot\alpha}_i &\to \tilde \lambda^{\dot\alpha}_i\, ,\quad \overline D^{\dot\beta}_i \overline{W}^{\dot\alpha}_i  \to \tilde \lambda^{\dot\beta}_i\tilde \lambda^{\dot\alpha}_i \eta_i, \quad D_{i,\alpha}\overline W^{\dot\alpha}_i = \lambda_{i,\beta}\frac{\partial}{\partial \eta_i} (\tilde \lambda^{\dot\alpha}_i) = 0
\end{align}
Note that, as in the chiral superfield case, the chirality conditions are upheld by the replacements. %due to the $U(N)$ covariance.

\begin{figure}[h!]
\begin{center}
\includegraphics[scale=0.3]{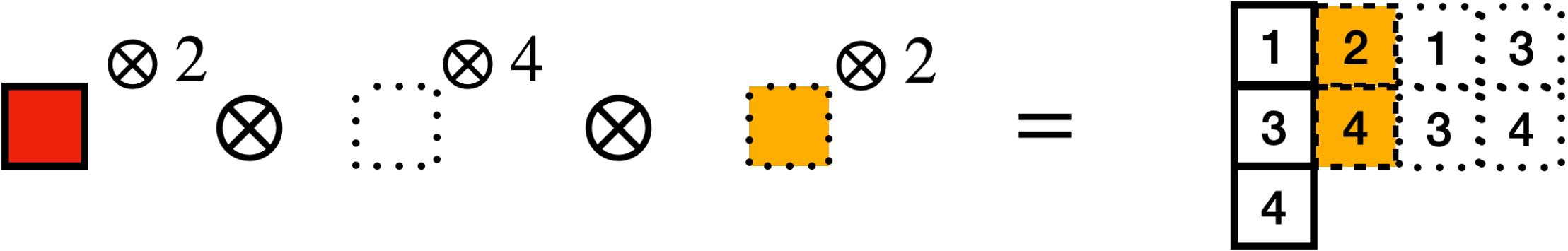}
\end{center}
\caption{Young tableau for the operator $(D\Phi_1)(\overline D\Phi^\dag_2)(DW_3) W_4 \overline W_5$, where all fields are massless and distinguishable. }
\label{vecexample}
\end{figure}

 We can now work out the YT for operators involving any combination of chiral/antichiral and vector superfields. Consider for example the operator class $D^2\overline D \Phi_1 \Phi^\dag_2 W_3 W_4 \overline W_5$, where $W_i$ are distinguishable abelian field strengths. Following the steps laid out in Sec.~\ref{sec:unsymm}, the first step is to determine the number of $\lambda$, $\tilde \lambda$, and $\eta$ present in the tensor product. As these numbers will be the same for all operators in the class, we can just pick a (legal) partitioning of the derivatives to find them, e.g.
\begin{align}
(D\Phi_1)(\overline D\Phi^\dag_2)(DW_3) W_4 \overline W_5  \to \lambda_1 (\tilde \lambda_2 \eta_2)( \lambda_3 \lambda_3)( \lambda_4 \eta_4) \tilde \lambda_5 \to n = 2, m = 4, N_c = 2
\end{align}
For this operator, there is a unique way we can contract the spinors: $\langle 2 5 \rangle [13][34]\eta_2\eta_4$. Applying steps 2.) - 4.) from Sec.~\ref{rules and examples} to this spinor helicity from, the filled-in YT is shown in Fig. \ref{vecexample}. 

In Table~\ref{dicv}, we summarize the updated replacement rule dictionary including vector fields. 

\begin{table}[h!]
\begin{center}
\begin{tabular}{ |c|c|}
 \hline
Off-shell & Spin-helicity Expression\\
 \hline
$\Phi_i $  & $\eta_i$   \\
$D\Phi_i $ & $\lambda_i$   \\
%$\overline{D}D\Phi_i $ & $\tilde{\lambda}_i\lambda_i\eta_i$   \\
$\Phi^\dag_i $  & $1$   \\
$\overline{D}\Phi^\dag_i $  & $\tilde{\lambda}_i\eta_i$   \\
$W_i$ & $\lambda_i \eta_i$ \\
$DW_i$ & $\lambda_i \lambda_i$ \\
$\overline W_i$ & $\tilde \lambda_i$ \\
$\overline D\overline W_i $ & $\tilde \lambda_i \tilde \lambda_i \eta_i$ \\
%${D\overline{D}\Phi^\dag_i} $    & $ (QQ^\dag\Phi^\dag)_i $&$\lambda_i\tilde{\lambda}_i$   \\
\hline
\end{tabular}
\end{center}
\caption{Replacement rule in the $\eta$ basis, expanded to include vector superfields. Adding additional powers of $D$ or $\overline D$ is straightforward and follows from the definitions in Eq.~\eqref{d rep}. }
\label{dicv}
\end{table}

Note that we have used $D_\alpha, \overline D^{\dot\alpha}$ to work out the YT, while in the actual operator construction these must of course be gauge covariant derivatives, $\nabla_\alpha, \overline{\nabla}^{\dot\alpha}$. This is analogous to the non-supersymmetric YT construction, where one treats $D_\mu \to \partial_\mu$ for the purposes of finding an operator's YT form. Additionally, going from YT to operator, we have no information about $e^{V_i}$ factors, as these carry gauge information only. Their correct placement must be worked out by hand by demanding gauge invariance once we know the operator's field and derivative content.
 
\section{Zero Derivative Term}\label{F term}
In this appendix, we show how to apply our approach to higher dimensional $F$-terms (superpotential terns) that contain no derivatives. Terms of this type are the product of chiral superfields only:
\begin{equation}
\mathcal{O}_F=\Phi_1\Phi_2\cdots\Phi_n.
\end{equation}
Applying our replacement rule and applying the $D^2$ from the $d^2\theta$ integration, we find:
\begin{equation}
A(\Phi_1\Phi_2\cdots\Phi_n)=\sum_{perm}^n \lambda_{a_1}\lambda_{a_2}\eta_{a_3}\cdots\eta_{a_n},
\end{equation}
where $(a_1,a_2, \cdots a_n)$ is the permutation of $(1,2, \dots, n)$.
Note that we don't include a $\delta^2(Q^\dag)$ piece here because an $F$-term only gets integrated over half of the superspace (e.g. $d^2\theta$ rather than $d^4\theta$). Despite the missing $\delta^2(Q^\dag)$, one can easily verify that this amplitude is annihilated by both $Q$ and $Q^\dag$. Annihilation by $Q$ follows via the Schouten identity exactly as in Eq.~\eqref{QZ}, while annihilation by $Q^\dag$ can be shown using momentum conservation $P\sim\sum_i \lambda_i\tilde{\lambda_i}\sim0$. Explicitly,
\begin{equation}
\begin{split}
Q^\dag A&=\sum_i (\tilde{\lambda}_i\eta_i) \sum_{perm}^n \lambda_{a_1}\lambda_{a_2}\eta_{a_3}\cdots\eta_{a_n}\\
&\sim \sum_{perm} (\tilde{\lambda}_{a_1}\eta_{a_1}+\tilde{\lambda}_{a_2}\eta_{a_2})\lambda_{a_1}\lambda_{a_2}\eta_{a_3}\cdots\eta_{a_n}\\
&\sim \sum_{perm} (-\tilde{\lambda}_{a_2}\lambda_{a_2}\lambda_{a_2}\eta_{a_1}+\tilde{\lambda}_{a_1}\lambda_{a_1}\lambda_{a_1}\eta_{a_2})\eta_{a_3}\cdots\eta_{a_n}\\
&\sim0
\end{split}
\end{equation}
where on the second line we use the fact that $\eta_i\eta_i=0$. The YT for such operators is a single column which has $n$ boxes and has the unique filling $(1,2,3,\cdots,n)^T$. 

\bibliographystyle{utphys}
\bibliography{ref}

\end{document}